\documentclass[twocolumn]{jpsj2} 
\renewcommand{\Vec}[1]{\mbox{\boldmath$#1$}}
\title{Pairing Symmetry Competition in Organic Superconductors}

\author{Kazuhiko \textsc{Kuroki}}

\inst{Department of Applied Physics and
Chemistry, The University of Electro-Communications,
Chofu, Tokyo 182-8585, Japan
}

\abst{A review is given on theoretical studies concerning the 
pairing symmetry in organic superconductors.
In particular, we focus on (TMTSF)$_2$X and $\kappa$-(BEDT-TTF)$_2$X, 
in which the pairing symmetry has been extensively 
studied both experimentally and 
theoretically. Possibilities of 
various pairing symmetry candidates and 
their possible microscopic origin are discussed. 
Also some tests for 
determining the actual pairing symmtery are surveyed.}

\kword{organic superconductivity, pairing symmetry, 
(TMTSF)$_2$X, $\kappa$-(BEDT-TTF)$_2$X, 
spin and charge fluctuations, Fermi surface}

\begin{document}
\maketitle

\section{Introduction}
\label{intro}
Possible occurrence of unconventional superconductivity 
in organic conductors\cite{ISY,ChemRev,OrgSupRev} 
has been of great interest recently.
Microscopically understanding 
the mechanism of pairing in those materials is 
an intriguing theoretical challenge.
Among the various candidates of unconventional superconductors, 
in this paper we will focus on two groups of superconductors 
in which the pairing symmetry has been extensively studied 
both theoretically and experimentally,  namely, 
(I) (TMTSF)$_2$X,\cite{TMTSFpairRev} 
or the Bechgaard salts, where TMTSF is an abbreviation for 
tetramethyltetraselenafulvalene and X stands for an anion such as 
PF$_6$, AsF$_6$, ClO$_4$, etc., and (II) $\kappa$-(BEDT-TTF)$_2$X,
\cite{kappaRev,kappaPairRev} 
where BEDT-TTF is an abbreviation for bisethylenedithio-tetrathiafulvalene 
and X=Cu(NCS)$_2$, Cu[N(CN)$_2$]Br, Cu$_2$(CN)$_3$, I$_3$, etc.
The key factors to be focused throughout the paper are 
the band structure and the shape of the 
Fermi surface, the band filling, and the wave number dependent pairing 
interactions mediated by spin and/or charge fluctuations and/or by phonons.
Superconductivity near charge ordered state as in 
$\theta$-(BEDT-TTF)$_2$X and $\alpha$-(BEDT-TTF)$_2$X\cite{Seo}
has also been investigated extensively, 
but will not be discussed here.\cite{Ogata}
Superconducting states induced under high magnetic fields,  
such as the Fulde-Ferrel-Larkin-
Ovchinnikov (FFLO) state\cite{FF,LO}, 
are also beyond the scope of the present paper.\cite{Uji2}


\section{(TMTSF)$_2$X}
\label{tmtsf}
\subsection{Lattice Structure and the Phase Diagram}
\label{latticetmtsf}
\begin{figure}
\begin{center}
\includegraphics[width=8cm,clip]{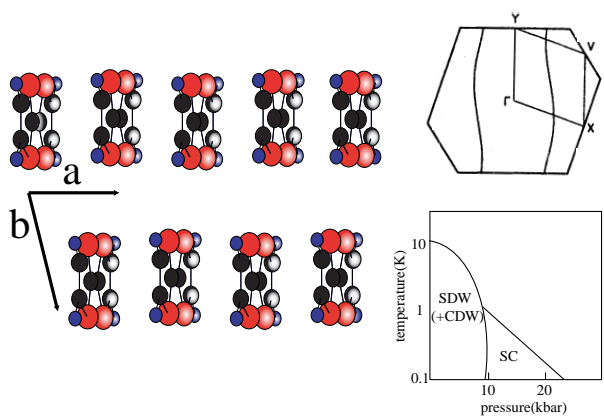}
\end{center}
\caption{Left panel:Lattice structure of (TMTSF)$_2$X in the $a$-$b$ plane.
Upper right panel: 
Typical shape of the Fermi surface.(Reprinted with permission from 
ref.\citen{Grant}. Copyright 1983 by EDP Sciences.) Lower right panel: 
Schematic phase diagram of (TMTSF)$_2$X.}
\label{fig1}
\end{figure}

The lattice structure of (TMTSF)$_2$X is shown in Fig.\ref{fig1}.
The molecules are stacked along the $a$-axis (denoted
as $\Vec{a}$ hereafter), which is the most conducting axis because the 
overlap of the molecular orbitals, oriented in the stacking direction, 
is large. The molecules are weakly dimerized along the stacks.
The charge transfer with the anions existing 
in between the conducting stacks results in  one hole 
per two molecules. 
There is a weak overlap of the orbitals in the $\Vec{b}$ 
direction, resulting in a weak two dimensionality.

A schematic phase diagram of (TMTSF)$_2$X is shown in Fig.\ref{fig1}.
At ambient pressure, (TMTSF)$_2$PF$_6$ undergoes a 
2$k_F$ spin density wave (SDW) transition at 12K. 
Upon increasing hydrostatic pressure, 
the SDW transition temperature decreases, and 
superconductivity with a transition temperature $(T_c)$ of $0.9$ K 
appears at $12$ kbar.\cite{Jerome} 
A similar phase diagram is obtained for X=AsF$_6$.\cite{Brusetti} 
It should be mentioned here that 
X-ray diffuse scattering experiments have revealed a coexistence of 
$2k_F$ charge density wave (CDW) in the SDW phase for X=PF$_6$,
\cite{Pouget,Kagoshima}
while the amplitude of the $2k_F$ CDW 
is very small for X=AsF$_6$.\cite{Kagoshima}
It should also be noted that the easy axis of the SDW is 
in the $\Vec{b}'$ direction,\cite{Mortensen,Mortensen2} 
which is the direction normal to the $\Vec{a}$-$\Vec{c}$ plane and 
somewhat tilted from $\Vec{b}$ due to the triclinic symmetry of the lattice.

(TMTSF)$_2$ClO$_4$ becomes superconducting at ambient pressure when 
the system is cooled down slowly enough for the anions to 
order at 24 K.\cite{Bechgaard}
On the other hand, when the cooling rate is fast, the anions are 
frozen in random directions, and in this case, SDW takes place 
instead of superconductivity.\cite{Takahashi}
It has also been revealed that superconductivity is destroyed 
upon alloying (TMTSF)$_2$ClO$_4$ with a small amount of ReO$_4$,
and with further alloying,  
an SDW phase appears.\cite{Coulon,Tomic,Joo1,Joo2}

\subsection{Electronic Structure}
\label{elecstructmtsf}
Reflecting the lattice structure and also the anisotropy of the 
orbitals, the band structure of (TMTSF)$_2$X is 
strongly one dimensional, i.e., the ratios of the hopping integrals in 
$\Vec{a}$, $\Vec{b}$, and $\Vec{c}$ directions are $t_b/t_a\sim 0.2$ 
and $t_c/t_b\sim 0.05$, where $t_a= 200\sim 300$ meV.\cite{Grant,Ducasse} 
Since $t_c$ is extremely small, it is highly likely that the 
essential mechanism of the superconductivity lies 
within the two dimensional lattice ($a$-$b$ plane). 
A typical Fermi surface is 
shown in Fig.\ref{fig1}, which is open in the $k_b$ direction due to the 
quasi-one-dimensionality.
We stress here that the 
anisotropy of the hopping integrals 
within the $a$-$b$ plane largely owes to 
the fact that molecular orbitals are directed toward the $\Vec{a}$ 
direction, while the distance between the molecules in the 
$\Vec{b}$ direction is only about two times larger than that in the 
$\Vec{a}$ direction. 
The hopping integral $t_a$ alternates along the $\Vec{a}$ direction 
by about $10\sim 20\%$  due to the dimerization of the molecules.
If we neglect this dimerization, the system is described by a 3/4-filled 
single band model, whose band dispersion is given as 
\begin{equation}
\varepsilon(\Vec{k})=2t_a\cos(k_a)+2t_b\cos(k_b).
\end{equation}
Here, only the hoppings between the nearest neighboring molecules in the 
$\Vec{a}$ and $\Vec{b}$ directions are considered. Lattice constants 
(neglecting the dimerization) are taken as the units of the length.
Many of the theoretical approaches have been based on this 
3/4-filled band model,  
but in some studies, the strong dimerization limit has been assumed, where 
each dimer of molecules is considered as a site, so that the band now becomes 
half filled.\cite{KinoKontani,NomuraYamada} There is also a study  
based on a two band model that maintains the realistic 
dimerization structure.\cite{KAA01}

\subsection{Experiments Concerning the Pairing Symmetry 
and Their Theoretical Interpretations} 
\label{exptmtsf}

Early experiments for (TMTSF)$_2$X, such as the specific heat 
\cite{Garoche,Brusetti2} and the upper critical field 
measurements\cite{Greene2,Chaikin,Murata} 
had been interpreted within the conventional $s$-wave pairing.
However, Abrikosov\cite{Abrikosov} 
pointed out the possibility of spin-triplet pairing 
based on the fact that $T_c$ is very sensitive to the 
existence of non-magnetic defects.\cite{Choi,Bouffard,Coulon,Tomic} 
More recently, Joo {\it et al.}\cite{Joo1,Joo2}  have shown that 
the sensitivity of $T_c$ to non-magnetic impurities (ReO$_4$) 
in (TMTSF)$_2$ClO$_4$ is precisely what is expected from 
the $T_c$ reduction formula\cite{AG,Larkin} for unconventional pairing.
Generally, 
in a superconducting state that satisfies the 
condition 
\begin{equation}
\sum_{\Vec{k}}F(\Vec{k},i\omega_n)=0,
\label{gapzero}
\end{equation}
the presence of non-magnetic impurities is pair breaking, and 
thus strongly suppresses $T_c$.\cite{SigristUeda} 
Here, $F$ is the anomalous Green's function, and the 
condition (\ref{gapzero}) roughly corresponds to 
a vanishing summation of the superconducting gap function 
$\Delta(\Vec{k})$ over the Fermi surface. 
The $T_c$ reduction in this case is given in the form,\cite{Larkin}
\begin{equation}
\ln\left(\frac{T_c}{T_{c0}}\right)
=\psi\left(\frac{1}{2}\right)-\psi\left(\frac{1}{2}+
\frac{\alpha}{2\pi T_c}\right),
\label{AGform}
\end{equation}
where $\psi$ is the digamma function, 
$T_{c0}$ is the transition temperature without impurities,
and $\alpha$ is the pair breaking parameter, which is determined by the 
scattering rate due to non-magnetic impurities.
This is in fact the same as the formula derived by 
Abrikosov and Gor'kov for the case of $s$-wave pairing with 
{\it magnetic} impurities.\cite{AG}
Since a triplet superconductivity has an odd parity gap,  
$T_c$ should be sensitive to the introduction of non-magnetic impurities.
Note, however, that the condition (\ref{gapzero}) can be satisfied 
for a superconducting state with an 
even parity gap that changes sign on the Fermi surface, 
so that the sensitivity 
to the presence of impurities alone of course does not 
necessarily imply triplet pairing. In this sense, the 
sensitivity of the $T_c$ to non-magnetic defects concerns the 
orbital part of the pair wave function.

Another experiment that indicated the possibility of 
unconventional pairing concerning the orbital part 
is the NMR experiment for X=ClO$_4$ 
performed by Takigawa {\it et al}.\cite{Takigawa}
Namely, the proton spin-lattice relaxation 
rate $1/T_1$ at zero magnetic field exhibits no coherence 
peak, and follows a power law 
temperature dependence close to $T^3$. 
Such a behavior is generally characteristic to superconductivity 
with a gap having line nodes.\cite{SigristUeda} In fact, 
Hasegawa and Fukuyama\cite{HaseFuku} 
studied various types of singlet and triplet anisotropic pairings
within the mean field approximation for a model with on-site and 
nearest neighbor attractive interactions, and calculated 
the spin-lattice relaxation rate. There it was shown 
that a singlet pairing 
without gap nodes on the Fermi surface exhibits a large coherence 
peak followed by an exponential decay of $1/T_1$, while for a triplet 
pairing with gap nodes at $k_a=0$ and thereby no nodes 
{\it on the Fermi surface}, which will be called 
$p$-wave hereafter (Fig.\ref{fig2}(c)), the coherence peak becomes smaller 
but still exists. Singlet and triplet pairings with line nodes of the
gap intersecting the Fermi surface 
cannot be distinguished from the temperature dependence 
of $1/T_1$; they both exhibit essentially no (or very small) coherence 
peak and a power law decay roughly proportional to $T^3$, 
which is similar to the experimentally observed behavior.\cite{Takigawa}
A more recent $1/T_1$ measurement has been performed on X=PF$_6$ 
by Lee {\it et al.},\cite{Lee1,Lee3} 
who have found a similar behavior of $1/T_1$ when a small 
magnetic field $\Vec{H}$ is applied parallel to  $\Vec{b'}$, 
but also an anomalous $1/T_1\sim T$ at low temperatures for high magnetic
fields.
On the other hand, Belin and Behnia showed for X=ClO$_4$ 
that the thermal conductivity rapidly decreases with lowering the 
temperature below $T_c$, indicating the absence of low lying 
excitations, and thus a fully gapped superconducting state.\cite{BB97}
A possible explanation for this discrepancy between the conclusions of 
the NMR and the thermal conductivity experiments 
will be discussed in section \ref{testtmtsf}.

Before discussing the experimental results  concerning the 
spin part of the pair wave function, 
let us briefly summarize some general aspects of spin triplet 
pairing.\cite{SigristUeda} 
In the case of triplet pairing, both the diagonal and the 
non-diagonal elements of the superconducting order parameter matrix
\begin{equation}
\hat{\Delta}(\Vec{k})=\left(
\begin{array}{cc} 
\Delta_{\uparrow\uparrow}(\Vec{k})&
\Delta_{\uparrow\downarrow}(\Vec{k})\\
\Delta_{\downarrow\uparrow}(\Vec{k})&
\Delta_{\downarrow\downarrow}(\Vec{k})
\end{array}\right)
\end{equation}
remain finite in general, where $\Delta_{\sigma\sigma'}$ are given as  
$\Delta_{\uparrow\uparrow}=-d_x+id_y$, 
$\Delta_{\uparrow\downarrow}=\Delta_{\downarrow\uparrow}=d_z$, 
and $\Delta_{\downarrow\downarrow}=d_x+id_y$, using the 
vector $\Vec{d}=(d_x,d_y,d_z)$. The order parameter vector $\Vec{d}$ 
lies in the direction perpendicular to the total spin of the triplet pairs.

Spin-triplet superconductivity can be identified by 
the NMR Knight shift measurement, which probes the uniform spin
susceptibility. The Knight shift decreases below $T_c$ for singlet 
pairing, while it stays constant for triplet pairing 
when the magnetic field $\Vec{H}$
is applied perpendicular to $\Vec{d}$.
Another possible way of detecting triplet pairing is to measure the 
upper critical field $H_{c2}$.
Cooper pairing under magnetic field 
is limited by both orbital and paramagnetic effects.
\cite{Clogston,Chandrasekhar} 
For triplet pairing, however, the paramagnetic limit (the Pauli limit, 
or the Clogston-Chandrasekhar limit) is overcome  when the 
magnetic field is applied perpendicular to $\Vec{d}$.

Now, as for the actual experimental results, 
Lee {\it et al.}  found for X=PF$_6$ that the 
Knight shift does not decrease below $T_c$ for magnetic fields 
applied parallel to  $\Vec{a}$\cite{Lee1} (1.43T) or  
$\Vec{b}'$ (2.38T) \cite{Lee3}. 
These results indicate that the 
pairing indeed occurs in the spin-triplet channel, and that 
either $\Vec{d}\parallel \Vec{c}$, or $\Vec{d}$ 
rotates in accord with the direction of the 
magnetic field to satisfy $\Vec{d}\perp\Vec{H}$.

As for the upper critical field $H_{c2}$, Gor'kov and J\'{e}rome pointed out 
in the early days that $H_{c2}$ extrapolated to $T=0$ 
may largely exceed the Pauli limit, suggesting the possibility of 
spin-triplet pairing.\cite{GorkovJerome} 
More recently, the upper critical field has been 
studied with higher accuracy and with  
precise orientation of the magnetic fields. 
$H^b_{c2}$, the upper critical field 
for  $\Vec{H}\parallel\Vec{b'}$ 
has been found to exceed the Pauli limit for 
X=PF$_6$\cite{Lee2} and also for ClO$_4$.\cite{Oh} 
Even if a spin-triplet pairing occurs, the pairing can still be 
orbitally limited, but  Lebed\cite{Lebed} and later 
Dupuis {\it et al.}\cite{DMS}  
showed that a magnetic field induced dimensional crossover from 
three to two dimensions can strongly enhance the 
orbital limit of the critical field. 
Thus, as far as $H_{c2}^b$ is concerned, the experimental results 
seem to be consistent with the above 
interpretations of triplet pairing 
with $\Vec{d}\parallel\Vec{c}$, or a rotatable $\Vec{d}$.\cite{commentFFLO}
However, the interpretation on the 
temperature dependence of $H_{c2}^a$ has been controversial.
For X=PF$_6$, there is an inversion between $H_{c2}^a$ and 
$H_{c2}^b$, where $H_{c2}^a>H_{c2}^b$  for small magnetic field, but 
$H_{c2}^a<H_{c2}^b$ for $H>1.6$T. Moreover, $H^a_{c2}(T)$ as a function of $T$ 
changes from a convex to a concave curve above $H=1.6$T.
From these experiments, Lebed {\it et al.} proposed that 
$d_b=0$ and $d_a\neq 0$ ($\Vec{d}=(d_a,d_b,d_c)$) 
assuming strong spin-orbit coupling, so that the  
pairing is Pauli-paramagnetically limited for $\Vec{H}\parallel\Vec{a}$ 
for $H<1.5$ T, while the change of the curvature of $H_{c2}^a(T)$ for higher
magnetic fields may be because 
$\Vec{d}$ rotates to become perpendicular to $\Vec{H}$, or may be due to 
an occurrence of the FFLO state.\cite{LMO}
On the other hand, Duncan {\it et al.} 
argued that spin-orbit coupling should be weak since the 
heaviest element in (TMTSF)$_2$X is Se, so that  $\Vec{d}$ 
should be able to rotate according to the direction of $\Vec{H}$ 
even for low magnetic fields.\cite{Duncan}
A clear understanding for the direction of $\Vec{d}$,
provided that triplet pairing does indeed occur,\cite{commentJerome} 
 requires further theoretical and experimental study.

\subsection{Spin-fluctuation-mediated $d$-wave pairing}
\label{dwavetmtsf}
In this and the next two subsections, we discuss some 
mechanisms for anisotropic pairing in  TMTSF salts.
Since the superconducting phase lies close to the SDW phase, 
and a number of experiments suggest the possibility of anisotropic, 
unconventional pairing, 
it is natural to expect that the spin fluctuations mediate  
(or at least play an important role in) the Cooper pairing 
in TMTSF salts, as was pointed out by Emery.\cite{Emery}
The spin-fluctuation-mediated pairing scenario has in fact 
been supported by several 
theoretical studies on the quasi-one-dimensional Hubbard model,
in which the on-site repulsive interaction $U$ is considered along with 
the kinetic energy part considered in section \ref{elecstructmtsf}.
The Hamiltonian is given in standard notation as, 
\[
H=\sum_{<i,j>,\sigma} 
t_{ij}c^{\dagger}_{i\sigma}c_{j\sigma}
+U\sum_{i}n_{i\uparrow}n_{i\downarrow},
\]
where $c^{\dagger}_{i\sigma}$ creates an electron with spin $\sigma$ at 
site $i$ (i.e., the $i$-th molecule), 
$n_{i\sigma}=c_{i\sigma}^\dagger c_{i\sigma}$, and 
$t_{ij}=t_a$ and $t_{ij}=t_b$ for intrachain and interchain nearest 
neighbor hoppings, respectively.
There, superconductivity has been studied using 
random phase approximation (RPA)\cite{Shimahara}, 
fluctuation exchange approximation (FLEX),\cite{KinoKontani}
third order perturbation,\cite{NomuraYamada} or 
quantum Monte Carlo method.\cite{KA} 
Here, based on RPA equations 
(which will be written down in a general form for later use) for the 
single band Hubbard model at quarter filling (quarter filling of holes, 
to be precise), 
we summarize the mechanism in which $2k_F$ spin fluctuations 
lead to $d$-wave like pairing.
Within RPA, singlet and triplet pairing interactions are 
given in the form,\cite{Scalapino,KobaSuzuOga,TanaOga}
\begin{eqnarray}
V^{s}(\Vec{q})&=&
U + V({\Vec q}) + \frac{3}{2}U^{2}\chi_{s}(\Vec{q})
-\frac{1}{2}(U + 2V({\Vec q}) )^{2}\chi_{c}(\Vec{q})\nonumber\\
V^{t}(\Vec{q})&=&
V({\Vec q}) - \frac{1}{2}U^{2}\chi_{s}(\Vec{q})
-\frac{1}{2}(U + 2V({\Vec q}) )^{2}\chi_{c}(\Vec{q}),
\label{pairint}
\end{eqnarray}
where $V(\Vec{q})$ is the Fourier transform of the 
off-site interactions (electron interactions between nearest neighbors, etc.),
which is 0 for the Hubbard model. 
%
Here, $\chi_{s}$ and $\chi_{c}$ are the spin and the charge 
susceptibilities, respectively,  which are given as 
\begin{eqnarray}
\label{4}
\chi_{s}(\Vec{q})=\frac{\chi_{0}(\Vec{q})}
{1 - U\chi_{0}(\Vec{q})},
\nonumber\\
\chi_{c}(\Vec{q})=\frac{\chi_{0}(\Vec{q})}
{1 + (U + 2V(\Vec{q}) )\chi_{0}(\Vec{q})}.
\label{RPA}
\end{eqnarray}
Here $\chi_{0}$ is the bare susceptibility given by 
\[
\chi_{0}(\Vec{q})
=\frac{1}{N}\sum_{\Vec{p}} 
\frac{ f(\varepsilon(\Vec{p +q}))-f(\varepsilon(\Vec{p})) }
{\varepsilon(\Vec{p}) -\varepsilon(\Vec{p+q})}
\]
where $f(\varepsilon)$ is the Fermi distribution function.
Within the weak coupling BCS theory, 
$T_c$ is obtained by solving  the linearized gap equation,
\begin{equation}
\lambda^{s,t} \Delta^{s,t}(\Vec{k})
=-\sum_{\Vec{k'}} V^{s,t}(\Vec{k-k'})
\frac{ \rm{tanh}(\beta \varepsilon(\Vec{k'})/2) }{2 \varepsilon(\Vec{k'}) }
\Delta^{s,t}(\Vec{k'}). 
\label{gapeq}
\end{equation}
The eigenfunction $\Delta^{s,t}$ of this eigenvalue equation is the 
gap function. 
The transition temperature $T_c$ is determined as the temperature 
where the eigenvalue $\lambda$ reaches unity. 
In the summation over $\Vec{k'}$ in the 
right hand side of eq.(\ref{gapeq}), the main contribution 
comes from $\Vec{k'}$ on the Fermi surface because of the factor 
$\frac{ \rm{tanh}(\beta \varepsilon(\Vec{k'})/2) }{2 \varepsilon(\Vec{k'}) }$.
If we multiply both sides of eq.(\ref{gapeq}) by 
$\Delta^{s,t}(\Vec{k})$ and take summation over the Fermi surface,
we see that the quantity
\begin{equation}
V^{s,t}_{\rm eff}=\frac{\sum_{\Vec{k},\Vec{k'}}V^{s,t}(\Vec{k}-\Vec{k'})
\Delta^{s,t}(\Vec{k})\Delta^{s,t}(\Vec{k'})}
{\sum_{\Vec{k}}(\Delta^{s,t}(\Vec{k}))^2}
\label{cond}
\end{equation}
has to be positive and 
large in order to have large $\lambda$, i.e., in order to 
have superconductivity with the gap $\Delta^{s,t}(\Vec{k})$.

\begin{figure}
\begin{center}
\includegraphics[width=8cm,clip]{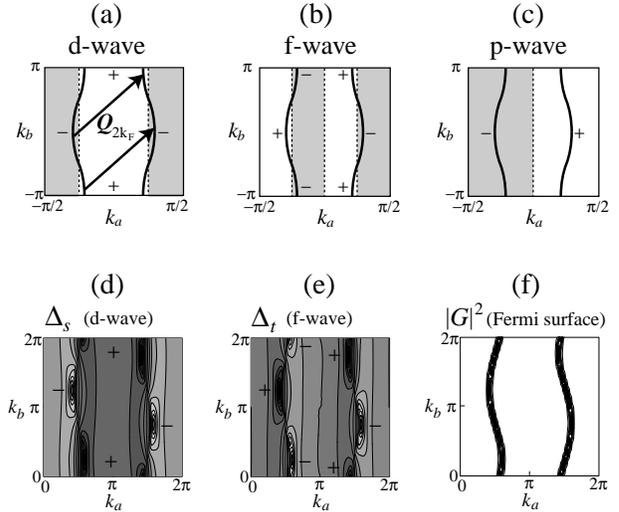}
\end{center}
\caption{Upper panel: candidates for the gap function of (TMTSF)$_2$X 
are schematically shown along with the Fermi surface (solid curves). 
(a)$d$-wave, (b)$f$-wave, (c)$p$-wave.
The dashed lines represent the nodes of the gap, 
whose $k_b$ dependence is omitted for simplicity. ``$+$,$-$'' represent the 
sign of the gap functions. Lower panel:FLEX 
calculation results for the two band model 
with finite dimerization.\cite{KAA01}
(d)$d$-wave gap, (e) $f$-wave gap, (f)$|G(\Vec{k},i\pi k_BT)|^2$, 
whose ridges represent the Fermi surface.}
\label{fig2}
\end{figure}

Due to the good nesting of the Fermi surface, 
the bare susceptibility $\chi_0(\Vec{q})$ peaks 
at the nesting vector $\Vec{q}=\Vec{Q}_{2k_F}$, and since 
$U>0$ and $V(\Vec{q})=0$,
$\chi_s(\Vec{q})$ becomes large at $\Vec{q}=\Vec{Q}_{2k_F}$,
while $\chi_c(\Vec{q})$ remains small at all $\Vec{q}$. 
Within this formulation, the SDW transition temperature is determined as the 
temperature where $U\chi_0(\Vec{Q}_{2k_F})$ reaches unity.
Thus, in the vicinity of the SDW transition, the 
pairing interactions roughly satisfy the relation
\begin{equation}
V^s(\Vec{Q}_{2k_F})=-3V^t(\Vec{Q}_{2k_F})>0.
\label{pairintrel}
\end{equation}
because the contribution from the spin fluctuations strongly 
dominates in eq.(\ref{pairint}).
Now, since the pairing interactions have large absolute values 
at $\Vec{q}=\Vec{Q}_{2k_F}$, the condition to have a positive 
$V_{\rm eff}$ in eq.(\ref{cond}) can be approximately reduced to 
\begin{equation}
V^{s,t}(\Vec{Q}_{2k_F})\Delta^{s,t}(\Vec{k})
\Delta^{s,t}(\Vec{k+Q}_{2k_F})<0,
\:\:\:\Vec{k},\Vec{k+Q}_{2k_F}\in {\rm F.S.}
\label{finalcond}
\end{equation}
From this condition and eq.(\ref{pairintrel}), 
we can see that the gap function 
has to change sign between $\Vec{k}$ and $\Vec{k+Q}_{2k_F}$ 
for singlet pairing, while the sign has to be the same across the 
nesting vector for triplet pairing. Since the spin part of the 
pair wave function is antisymmetric (symmetric) with respect 
to the exchange of electrons for spin singlet (triplet) pairing,
the orbital part of the wave function, namely the gap function,
has to satisfy the condition $\Delta^s(\Vec{k})=\Delta^s(-\Vec{k})$ 
(even parity gap)
and $\Delta^t(\Vec{k})=-\Delta^t(-\Vec{k})$ (odd parity), 
for singlet and triplet 
pairings, respectively. The gap functions satisfying these conditions 
are schematically shown in Fig.\ref{fig2}(a)(b).
We will call the singlet pairing ``$d$-wave'' in the sense that the 
gap changes sign as $+-+-$ {\it along the Fermi surface}, while the 
triplet pairing will be called ``$f$-wave'' in the sense that the 
gap changes sign as $+-+-+-$.\cite{symmetrycomment} 
Since the pairing interaction is 
three times larger for the singlet pairing, $d$-wave pairing 
takes place in this case. Note that a simpler form of an  
odd parity gap is the $p$-wave shown in Fig.\ref{fig2}(c), which 
changes sign as $+-$ along the Fermi surface. However, 
this gap does not satisfy the condition (\ref{finalcond}) 
because the triplet pairing interaction is negative for the 
Hubbard model at least within RPA.

Although we have adopted RPA equations in the 
above, similar conclusions have been drawn from 
other approaches as mentioned above. 
For example, an approach along the line of RPA, but more 
suitable for dealing with strong spin fluctuations, 
is the FLEX method.\cite{Bickers} 
In the FLEX, 
(i) Dyson's equation is 
solved to obtain the renormalized Green's function $G(k)$,
where $k\equiv({\bf k},i\epsilon_n)$ denotes the wave vectors and 
the Matsubara frequencies,
(ii) the effective electron-electron interaction $V^{(1)}(q)$ 
is calculated by collecting RPA-type  diagrams consisting
of the renormalized Green's function, namely, 
by summing up powers of the irreducible susceptibility 
$\chi_{\rm irr}(q)\equiv -\frac{1}{N}\sum_k G(k+q)G(k)$ 
($N$:number of $k$-point meshes),
(iii) the self energy is obtained as 
$\Sigma(k)\equiv\frac{1}{N}\sum_{q} G(k-q)V^{(1)}(q)$, 
which is substituted into Dyson's equation in (i), 
and the self-consistent loops are repeated until convergence is attained.

To obtain $T_c$, the linearized 
{\'E}liashberg equation for the singlet or the triplet gap function 
$\Delta^{s,t}(k)$, 
\begin{equation}
\lambda\Delta^{s,t}(k)=-\frac{T}{N}
\sum_{k'}
 V^{s,t}(k-k')G(k')G(-k')\Delta^{s,t}(k'),
\label{eliash}
\end{equation}
is solved, 
where the singlet or the triplet pairing interactions $V^{s,t}$ are given 
again in the RPA form but using the irreducible susceptibility 
 obtained from the renormalized Green's functions instead of the bare 
susceptibility. $T_c$ is the temperature 
where the eigenvalue $\lambda$ reaches unity.

Kino and Kontani applied FLEX to the half-filled Hubbard model,
i.e., the model in the strong dimerization limit,\cite{KinoKontani} 
where they obtained a finite $T_c$ for the $d$-wave pairing.
Kuroki {\it et al.} applied FLEX to a two-band model 
with finite dimerization and with next nearest neighbor 
interchain hoppings, and also found that the 
$d$-wave pairing (Fig.\ref{fig2}(d)) is the most dominant pairing, 
while triplet $f$-wave pairing (Fig.\ref{fig2}(e)) is subdominant.
\cite{KAA01}

Another approach for the Hubbard model is 
the perturbational theory, where all the Feynman diagrams up to 
a certain order are taken into account in the calculation of 
the pairing interactions.
Applying the third order perturbation theory to the half-filled 
model in the dimer limit, Nomura and Yamada obtained finite values of 
$T_c$ for the 
$d$-wave pairing. It has been found there also that 
the $f$-wave pairing is subdominant.
\cite{NomuraYamada}

As for numerical calculations for finite size systems, 
Kuroki and Aoki\cite{KA} adopted 
the ground state quantum Monte Carlo (QMC) technique
\cite{Sorella,White,Imada}. 
This method enables us to accurately calculate 
correlation functions within statistical errors for finite size 
clusters. Since the superconducting order parameter is always zero 
for finite size systems, we instead calculate its fluctuation, namely, 
the pairing correlation 
function, given in the form 
$\langle c_{i+\delta}c_{i}c^\dagger_{j}c^\dagger_{j+\delta}\rangle$, 
where $i,j$ denotes the sites, and $i$ and $i+\delta$ are the sites 
at which the Cooper pair is formed. When the tendency towards 
superconductivity is strong, the pairing correlation decays slowly at 
large distances between sites $i$ and $j$.
Applying this method to the single band Hubbard model at quarter filling,
it has been found that the $d$-wave pairing correlation function is 
enhanced at large distances by the presence of the on-site repulsion
$U$.\cite{KA} More recently, Kuroki {\it et al.} studied the 
pairing symmetry competition on the Hubbard model at quarter filling using the 
ground state QMC, where they found that $d$-wave and $f$-wave 
strongly dominate over $p$-wave.\cite{KTKA}

Apart from the studies directly dealing with the Hubbard model, 
low energy theories using the interacting 
electron gas model like those for the 
purely one dimensional systems as will be mentioned 
in section \ref{phonontmtsf} can be effective, but 
since the nodes of the $d$-wave gap run parallel to the $k_b$ axis, 
it is necessary to take into account the 
quasi-one-dimensionality (the warping of the Fermi surface) to study 
$d$-wave pairing in a realistic situation. Duprat and Bourbonnais indeed 
showed the occurrence of $d$-wave 
pairing near the SDW phase within a renormalization 
group study that takes into account the quasi one dimensionality.\cite{Duprat}

\subsection{Spin triplet $f$-wave pairing}
\label{fwavetmtsf}
Nevertheless, the spin-fluctuation-mediated $d$-wave pairing scenario 
contradicts with the experimental facts pointing towards 
spin-triplet pairing, especially for X=PF$_6$.\cite{Lee1,Lee2} 
(Note that most of the $d$-wave theories appeared before the 
Knight shift measurements.)
Kuroki {\it et al.}\cite{KAA01} provided a possible solution for this 
puzzle by recalling that $2k_F$ CDW actually coexists with 
$2k_F$ SDW in the insulating phase for X=PF$_6$.\cite{Pouget,Kagoshima}
If $2k_F$ CDW coexists with SDW in the insulating phase, 
it is natural to assume that $2k_F$ spin and $2k_F$ charge fluctuations 
coexist in the metallic phase lying nearby. Assuming the presence of 
charge fluctuations along with spin fluctuations with possible 
magnetic anisotropy (i.e.,presence of easy and hard axes), 
the pairing interactions are given in generic forms, 
\begin{eqnarray}
V^s(\Vec{q})&=&\frac{1}{2}V^{zz}_{\rm sp}(\Vec{q})+V^{+-}_{\rm sp}(\Vec{q})
-\frac{1}{2}V_{\rm ch}(\Vec{q})\nonumber\\
V^{t\perp}(\Vec{q})&=&
-\frac{1}{2}V^{zz}_{\rm sp}(\Vec{q})-\frac{1}{2}V_{\rm ch}(\Vec{q})\nonumber\\
V^{t\parallel}(\Vec{q})&=&
\frac{1}{2}V^{zz}_{\rm sp}(\Vec{q})-V^{+-}_{\rm sp}(\Vec{q})
-\frac{1}{2}V_{\rm ch}(\Vec{q})
\label{pairt0}
\end{eqnarray}
where $V^{zz}_{\rm sp}$ and $V^{+-}_{\rm sp}$ are the contributions 
from longitudinal and transverse spin fluctuations, respectively, 
while $V_{\rm ch}$ is the contribution from the charge fluctuations.
There are two triplet pairing interactions:
$V^{t\perp}$ for $\Vec{d}\perp\Vec{z}$ and $V^{t\parallel}$ for 
$\Vec{d}\parallel\Vec{z}$.
The contribution from the spin fluctuations is expected to be large in the 
easy axis direction of the SDW ordering. Then, taking the easy axis as the 
$z$-axis, we may assume $V^{zz}_{\rm sp}(\Vec{Q}_{2k_F})
\geq V^{+-}_{\rm sp}(\Vec{Q}_{2k_F})$
because the longitudinal spin susceptibility should exhibit stronger 
divergence at $\Vec{q}=\Vec{Q}_{2k_F}$ than the transverse ones 
near the SDW transition. Thus,  
$-V^{t\perp}(\Vec{Q}_{2k_F})\geq -V^{t\parallel}(\Vec{Q}_{2k_F})$ holds 
from eq.(\ref{pairt0}), where $-V^{t\perp}(\Vec{Q}_{2k_F})$ is always positive.
Furthermore from eq.(\ref{pairt0}), we can see that 
\begin{equation}
-V^{t\perp}(\Vec{Q}_{2k_F})\geq V^s(\Vec{Q}_{2k_F})
\end{equation}
holds when the condition,
\begin{equation}
V_{\rm ch}(\Vec{Q}_{2k_F})\geq V^{+-}_{\rm sp}(\Vec{Q}_{2k_F})
\label{condition1}
\end{equation}
is satisfied. This kind of relation between the singlet and the triplet 
pairing interactions when spin and charge fluctuations coexist 
has been pointed out by Takimoto\cite{Takimoto} for another 
candidate for a triplet superconductor, Sr$_2$RuO$_4$.\cite{Maeno}

Now, an important point for a quasi-one-dimensional system 
is that the number of gap nodes that intersect the Fermi 
surface is the same between $d$- and $f$-waves due to the 
disconnectivity of the Fermi surface, so that which one of 
these two dominates is determined solely by the magnitude of 
the pairing interactions. Thus, triplet $f$-wave pairing with 
$\Vec{d}$ perpendicular to the easy axis direction  
dominates over singlet $d$-wave when the contributions to the 
pairing interaction from the charge fluctuations is 
larger than that from the spin fluctuations 
in the hard axis direction.

\begin{figure}
\begin{center}
\includegraphics[width=8cm,clip]{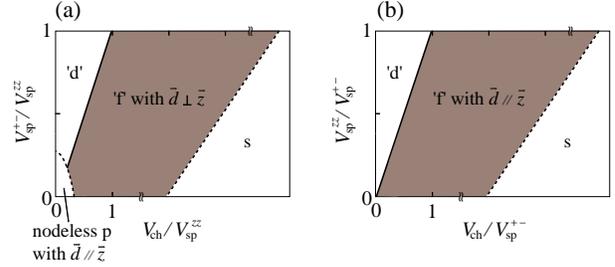}
\end{center}
\caption{Phenomenological phase diagram of the pairing symmetry 
for (a)$\Vec{z}\parallel\Vec{b}$\cite{KAA01} 
and (b) $\Vec{z}\parallel\Vec{c}$.
}
\label{fig3}
\end{figure}

The above argument can be summed up as a phenomenological 
phase diagram shown in Fig.\ref{fig3}(a).
In this phase diagram, there exists a region where 
$p$-wave pairing dominates because 
$V^{t\parallel}(\Vec{Q}_{2k_F})>0$ holds 
when $V^{zz}_{\rm sp}>2V^{+-}_{\rm sp}+V_{\rm ch}$,
namely, when the magnetic anisotropy is strong and the charge 
fluctuations are weak, so that a triplet gap that 
has different signs at both ends of the nesting vector 
can be favored. In this case,
$\Vec{d}$ is parallel to the easy axis. 
This $p$-wave mechanism has in fact been proposed for Sr$_2$RuO$_4$.
\cite{KuwaOga,SatoKoh} $s$-wave pairing having the same gap sign over the 
entire Fermi surface is expected to dominate 
when the charge fluctuations are sufficiently strong because 
the singlet pairing interaction turns negative (which 
is unrealistic for (TMTSF)$_2$X).

\begin{figure}
\begin{center}
\includegraphics[width=8cm,clip]{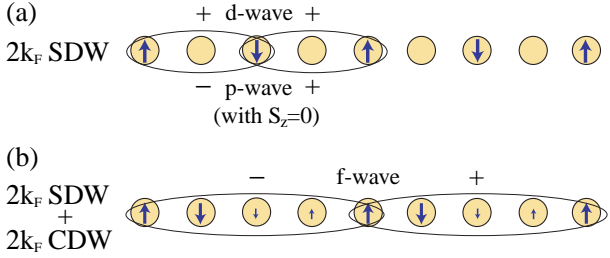}
\end{center}
\caption{(a)$2k_F$ SDW configuration and (b) $2k_F$ SDW+$2k_F$ CDW
 configuration with likely pairings when the configuration ``melts'' to 
become metallic.
}
\label{fig4}
\end{figure}

An intuitive understanding for this phase diagram can be given as follows.
In the $2k_F$ SDW configuration, electrons 
(or, actually, holes in a 3/4-filled system) 
with antiparallel spins 
sit at next nearest neighbors as shown in Fig.\ref{fig4}(a), so if this 
configuration ``melts'' to become metallic, singlet pairing 
superconductivity with an  even parity gap of 
$\Delta(\Vec{k})=+\exp(i2k_a)+\exp(-i2k_a)\sim\cos(2k_a)$ 
is likely to occur. ``2'' in the argument of ``$\exp$'' implies that the pairs 
are formed at second nearest neighbors,
and the ``+'' signs in front of the ``$\exp$'' 
corresponds to singlet wave functions  
having the same sign in the right and the left directions, 
as shown in Fig.\ref{fig4}(a). Since the gap $\cos(2k_a)$ has even parity and 
has nodes at $k_a=\pm \pi/4$, 
this corresponds to the singlet $d$-wave. On the other hand, 
when $2k_F$ SDW and $2k_F$ CDW coexist (namely when both SDW and CDW 
have a period of four lattice spacings), the electrons are aligned 
like in Fig.\ref{fig4}(b), so that when this configuration melts, 
triplet superconductivity with an odd parity gap of 
$\Delta(\Vec{k})=+\exp(i4k_a)-\exp(-i4k_a)\sim\sin(4k_a)$ 
is likely to take place.
This corresponds to the $f$-wave gap. If we consider the magnetic 
anisotropy, a triplet pair formed at fourth nearest neighbors 
is expected to have a total $S_z=\pm 1$  
because the SDW spins are oriented in the $\Vec{z}$ direction, 
($\Vec{z}\parallel\Vec{b}'$), which explains  $\Vec{d}\perp\Vec{z}$ for 
$f$-wave.
On the other hand, if the pure $2k_F$ SDW configuration (Fig.\ref{fig4}(a)) 
with $\Vec{z}$ being the easy axis melts, 
a triplet pairing with $S_z=0$ formed at 
next nearest neighbor sites may compete with the singlet pairing.
This corresponds to the $p$-wave pairing with the gap 
$\Delta(\Vec{k})=+\exp(i2k_a)-\exp(-i2k_a)\sim\sin(2k_a)$, whose 
nodes do not intersect the Fermi surface. Since the total spin 
of a triplet pair in this case is expected to be perpendicular to 
$\Vec{z}$, $\Vec{d}\parallel\Vec{z}$ can be understood.

In the above, the $z$-axis of the spins is taken in the 
$\Vec{b}'$ direction, namely, the easy axis direction, 
but if we assume that the spin fluctuations in the 
$a$-$b'$ plane are nearly isotropic and larger than those in the 
$\Vec{c}$ (hard axis) direction, we can take
the hard axis as the $z$-axis and thus 
$V_{\rm sp}^{+-}(\Vec{Q}_{2k_F})>V_{\rm sp}^{zz}(\Vec{Q}_{2k_F})$,
so that now $\Vec{d}\parallel\Vec{c}$ for $f$-wave pairing following 
a similar argument as before. 
This picture may be 
more suitable for (TMTSF)$_2$PF$_6$ since (i) the uniform susceptibilities 
for $\Vec{H}\parallel\Vec{a}$ and $\Vec{H}\parallel\Vec{b}'$ 
are equal down to the very 
vicinity of the SDW transition, while that for $\Vec{H}\parallel\Vec{c}$ 
deviates from higher temperatures,
and (ii) the direction of the 
SDW undergoes a spin-flop transition into the $\Vec{a}$ direction 
under a magnetic field in the $\Vec{b}'$ direction,\cite{Mortensen,Mortensen2} 
which may be an indication that the $2k_F$ spin  
fluctuations in the metallic state may not be 
so anisotropic within the $a$-$b'$ plane.
The phase diagram for $\Vec{z}\parallel\Vec{c}$ assuming 
isotropic spin fluctuations in the $a$-$b'$ plane is shown in 
Fig.\ref{fig3}(b).\cite{Aizawa} Note that in this case, 
strong anisotropy in the spin fluctuations 
does not lead to $p$-wave pairing because the triplet pairing 
interactions always remain negative.

\begin{figure}
\begin{center}
\includegraphics[width=8cm,clip]{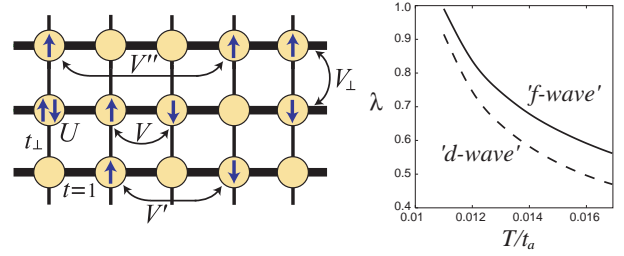}
\end{center}
\caption{Left: The model for (TMTSF)$_2$X adopted in ref.
\citen{KurokiTanaka}. Right: The largest eigenvalue of the gap equation 
in the singlet and the triplet 
channels are plotted as functions of temperature for 
$U=1.7$, $V=0.8$, $V'=0.45$, $V''=0.2$, $V_\perp=0.4$, $t_b=0.2$, in units 
of $t_a$. 
\cite{KurokiTanaka}
}
\label{fig5}
\end{figure}

After this phenomenological proposal and also a similar 
phenomenological argument of $f$-wave pairing by Fuseya {\it et 
al.},\cite{Fuseya1} studies based on microscopic models 
have followed.  Tanaka and Kuroki considered a model 
which takes into account the off-site repulsive interactions 
$\sum_{<i,j>}V_{ij} n_{i}n_{j}$ within the chains 
up to third nearest neighbors (Fig.\ref{fig5}, but with $V_\perp=0$), 
where the consideration of the second nearest neighbor repulsion 
$V'$ is the key.\cite{TanakaKuroki04} 
This has been based on a consideration that  
since the coexistence of $2k_F$ spin and 
$2k_F$ charge fluctuations is 
necessary for $f$-wave pairing, 
and since the coexistence $2k_F$ SDW and CDW is 
experimentally observed,\cite{Pouget,Kagoshima}
a model that can account for this coexistence 
should be the right Hamiltonian to be adopted. 
The mechanism of the coexistence of $2k_F$ SDW and $2k_F$ CDW itself 
had already been proposed by Kobayashi {\it et al.} 
and also studied by Tomio and Suzumura,
where the second nearest neighbor repulsion $V'$ plays an essential
role.\cite{KobaOga,KobaOga2,Suzumura,Ramasesha}
From  Fig.\ref{fig4}, it can be seen how $V'$ induces the $2k_F$ CDW in a 
quarter-filled system.
When only the on-site $U$ and the nearest neighbor $V$ are present,
the charges tend to take the $4k_F(=\pi)$ CDW configuration, 
which has a period of two sites, while when $V'$ is present, 
the pairs of electrons sitting at second neighbors repel each other to 
result in the $2k_F$ CDW(+SDW) configuration.

For the Hamiltonian that considers $U$, $V$, $V'$, and $V''$,
the Fourier transform of the off-site repulsions, 
considered in the RPA eq.(\ref{RPA}),  is given as 
\begin{equation}
V(\Vec{q}) =2V\cos(q_x)+2V'\cos(2q_x)+2V''\cos(3q_x)
\label{fourier}
\end{equation}
From eqs.(\ref{pairint}),(\ref{RPA}), and (\ref{fourier}), 
it can be seen that $\chi_s(\Vec{Q}_{2k_F})=\chi_c(\Vec{Q}_{2k_F})$
(where $\Vec{Q}_{2k_F}=(\pi/2,\pi)$), and consequently 
$-V^t(\Vec{Q}_{2k_F})=V^s(\Vec{Q}_{2k_F})$, apart from the first 
order terms such as $U+V(\Vec{q})$, is satisfied when $V'=U/2$.
Within the phenomenological argument, 
this corresponds to the condition for $f$-wave to be degenerate with 
$d$-wave in the absence of magnetic anisotropy.\cite{commentmaganiso} 
In the actual RPA
calculation for $V'=U/2$, $f$-wave slightly dominates over $d$-wave due to the 
effect of the first order terms neglected in the phenomenological 
argument.\cite{TanakaKuroki04}

Fuseya and Suzumura\cite{Fuseya04} 
approached the same problem using the renormalization 
group method for quasi-one-dimensional systems 
along the line of Duprat and Bourbonnais\cite{Duprat}, 
where a similar conclusion has been reached. 
Since the pairing competition is subtle,
they further proposed a possible singlet 
$d$-wave to triplet $f$-wave transition in the presence of magnetic field. 

According to the above studies, 
$f$-wave dominates over $d$-wave when the 
second nearest neighbor repulsion $V'$ is equal to or larger than 
half the on-site repulsion $U$, which may be 
difficult to realize in actual materials.
Nickel {\it et al.} have proposed a possible solution for this 
difficulty, where they considered, in addition to the 
intrachain repulsions, the {\it interchain} repulsion
and used the renormalization group technique 
for quasi-one-dimensional systems to reach 
a conclusion that $f$-wave 
dominates over $d$-wave in a more realistic parameter regime with 
a smaller second nearest neighbor repulsion.\cite{TMTSFpairRev,Nickel}
Independently, Kuroki and Tanaka also considered a model 
that considers the nearest neighbor interchain 
repulsion $V_\perp$ as shown in Fig.\ref{fig5}.\cite{KurokiTanaka} 
Within RPA, the term 
$2V_{\perp}\cos(q_y)$ is added in the right hand side of eq.(\ref{fourier}),
so that the condition for
$\chi_s(\Vec{Q}_{2k_F})=\chi_c(\Vec{Q}_{2k_F})$ now becomes
$V'+V_\perp=U/2$. This is a much more realistic condition than 
$V'=U/2$ because the interchain distance is similar to the 
intrachain second nearest neighbor distance, so that we can 
expect $V_\perp$ to be as large as $V'$. The actual 
calculation shows that $f$-wave dominates (has a larger eigenvalue $\lambda$) 
over $d$-wave for a parameter set, e.g.,  
$U=1.7$, $V=0.8$, $V'=0.45$, $V''=0.2$, $V_\perp=0.4$, $t_b=0.2$ 
in units of $t_a$(Fig.\ref{fig5}), 
where the relative magnitude of the 
interactions can be considered as realistic.

\subsection{Other Mechanisms for Triplet Pairing: Phonons, 
Ring Exchange}
\label{phonontmtsf}
In this subsection, we discuss some other mechanisms for spin-triplet 
pairing proposed for (TMTSF)$_2$X. From the early days, possibility of 
spin-triplet superconductivity in (TMTSF)$_2$X has been 
discussed in terms of the low energy effective theory called  the 
$g$-ology approach for the purely one dimensional interacting 
electron gas, i.e., the Tomonaga-Luttinger model.\cite{TLRev}
In the $g$-ology phase diagram, the spin-triplet superconducting 
phase and the SDW phase share boundary,\cite{commentgology} 
so that it is tempting to relate this 
superconducting state with that of (TMTSF)$_2$X, as was discussed in 
some studies.\cite{Barisic1}  
More recently, this phase boundary between the SDW and the 
triplet superconductivity has been discussed as having SO(4) 
symmetry.\cite{Rozhkov,Podolsky} 
Nevertheless, since exact numerical studies 
on the {\it purely} one-dimensional extended Hubbard model, 
where the on-site $U$ and the 
nearest neighbor $V$ is considered,\cite{SanoOno} show that 
superconductivity does not occur in a realistic parameter regime 
when the interactions are all repulsive, it is likely that 
some kind of attractive interaction, most probably originating from 
electron-phonon interaction, should be necessary in order to realize the 
triplet superconducting state 
in the $g$-ology phase diagram, as discussed in some 
studies\cite{Horovitz,Barisic2}. 

Apart from the $g$-ology-type approach, there have been 
studies on the phonon mechanism of triplet pairing.
Kohmoto and Sato proposed  a $p$-wave pairing mechanism 
due to a combination of 
electron-phonon interaction, 
$2k_F$ spin fluctuations, and the 
disconnected Fermi surface.\cite{KohSato}  Assuming that the 
electron-phonon interaction is weakly screened, a long-ranged attractive
interaction arises in real space, which means that 
the pairing interaction becomes large and negative 
around $\Vec{q}\sim0$ in momentum space. If we denote this interaction as
$-V_{\rm el-ph}(\Vec{q})$, and if the spin fluctuations 
also contribute to the pairing to some extent, 
the pairing interactions are given as 
\begin{eqnarray}
V^s(\Vec{q})&=&-V_{\rm el-ph}(\Vec{q})+
\frac{1}{2}V^{zz}_{\rm sp}(\Vec{q})+V^{+-}_{\rm sp}(\Vec{q})\nonumber\\
V^{t\perp}(\Vec{q})&=&-V_{\rm el-ph}(\Vec{q})
-\frac{1}{2}V^{zz}_{\rm sp}(\Vec{q})\nonumber\\
V^{t\parallel}(\Vec{q})&=&-V_{\rm el-ph}(\Vec{q})+
\frac{1}{2}V^{zz}_{\rm sp}(\Vec{q})-V^{+-}_{\rm sp}(\Vec{q}).
\label{phononspin}
\end{eqnarray}
In ref.\citen{KohSato}, the competition between $s$- and 
$p$-wave pairings was discussed, while 
the possibility of $d$- and $f$-wave was 
not considered because the warping of the Fermi surface was neglected.
Let us first neglect the magnetic anisotropy, i.e., 
$V^{zz}=V^{+-}$.
Around $\Vec{q}\sim 0$, neglecting the spin fluctuation contribution, 
the pairing interaction is negative  
and has the same magnitude between singlet and triplet pairings. 
Thus, $s$- and $p$-wave pairings, whose gap does not change sign on 
each portion of the disconnected Fermi surface, are equally favored by 
this interaction around $\Vec{q}\sim 0$. 
At $\Vec{q}=\Vec{Q}_{2k_F}$ on the other hand, 
neglecting the electron-phonon interaction  
this time, the positive spin fluctuation contribution for the singlet 
pairing works destructively 
against $s$-wave because the gap does not 
change sign across $\Vec{Q}_{2k_F}$, while the negative contribution for the 
triplet channel also works against $p$-wave, whose gap changes sign. 
Since this destructive spin fluctuation contribution is 
smaller for triplet pairing, 
$p$-wave dominates over $s$-wave. Note that here again, the 
close competition between $p$-wave and $s$-wave arises from the 
disconnectivity of the Fermi surface owing to the (quasi) one 
dimensionality, i.e, the additional node in the $p$-wave gap as 
compared to the $s$-wave does not intersect the Fermi surface.

If we further take into account the magnetic anisotropy and assume 
$V^{zz}<V^{+-}$ by taking the hard axis ($c$-axis) in the $z$ direction, 
the negative spin fluctuation contribution in the triplet pairing interaction 
is smaller (and thus favorable for $p$-wave pairing)
for $\Vec{d}\perp\Vec{z}$ than for $\Vec{d}\parallel\Vec{z}$. 
Therefore, if the direction of $\Vec{d}$ for $p$-wave pairing is 
governed by the magnetic anisotropy of the SDW, 
$\Vec{d}$ is likely to be perpendicular to the hard axis direction,
namely, in the $a$-$b$ plane for (TMTSF)$_2$PF$_6$. 

Suginishi and Shimahara also proposed a phonon-mediated mechanism for 
$p$-wave pairing.\cite{SugiShima} 
By considering moderately screened phonons and also 
including the corrections due to 
charge fluctuations,  
they obtained an attractive pairing interaction that 
has a large magnitude around $\Vec{q}\sim 0$ and a small one around 
$\Vec{q}\sim\Vec{Q}_{2k_F}$. By further considering the 
Coulomb pseudo potential, which suppresses only the $s$-wave 
pairing, it has been found there that $p$-wave pairing dominates in a certain 
parameter regime.

Recently, Ohta {\it et al.} proposed a non-electron-phonon 
mechanism for spin-triplet pairing.\cite{Ohta} 
The mechanism is based on the fact that in a triangle lattice 
consisting of three sites with two electrons, 
a ferromagnetic interaction arises by considering a consecutive 
exchange of the positions of the electrons.\cite{Penc}
If ferromagnetic spin fluctuations arise due to this 
``ring exchange mechanism'' on a certain lattice, 
triplet pairing superconductivity may take place. 
They considered the Hubbard model on a one-dimensional 
``railway-trestle'' (or zigzag) lattice,
where they used the density matrix renormalization group method 
to find that triplet pairing correlation functions decay
more slowly than the singlet ones. Their numerical calculation
has been restricted to purely one dimensional systems so far, but they 
further propose that 
this mechanism may be applicable to the quasi-one-dimensional 
material (TMTSF)$_2$X since the signs of the intrachain and interchain 
hopping integrals (by considering also the next nearest neighbor 
interchain hopping)\cite{Grant,Ducasse}  satisfy the condition
for the ferromagnetic interaction.\cite{Penc} 

In all the mechanisms discussed in this subsection, at least one of the 
$2k_F$ fluctuations, spin or charge, are not taken into account, 
although they should both be present at least for X=PF$_6$.
Then, whether both of these fluctuations play essential roles or not 
in the occurrence of superconductivity 
is the key toward clarifying 
whether $f$-wave discussed in section \ref{fwavetmtsf} 
or other triplet pairings  dominate, 
provided that the pairing indeed occurs in the triplet channel.

\subsection{Tests for the Pairing Symmetry Candidates}
\label{testtmtsf}
In this section, we discuss some experimental tests (already existing
ones as well as proposals for future study) for the 
candidates for the pairing symmetry discussed above.
For the pairing symmetries whose gap has line nodes on the Fermi surface 
such as $d$-wave and  $f$-wave, the spin-lattice relation rate $1/T_1$ 
exhibits essentially no (or very small) coherence peak and a 
power law decay proportional to $\sim T^3$,\cite{HaseFukucomment} 
which is consistent with the 
experiments for X=ClO$_4$\cite{Takigawa} and for X=PF$_6$\cite{Lee3} 
at low magnetic fields.
On the other hand, whether these pairings can account for the peculiar 
behavior of $1/T_1$ observed for X=PF$_6$ at high magnetic field, i.e., 
$1/T_1\sim T$ at low temperatures as well as a small peak below $T_c$ for 
$\Vec{H}\parallel\Vec{a}$,\cite{Lee1,Lee3} 
remains open as an interesting future study. 

At first glance, only $p$-wave and $s$-wave pairings seem to be 
consistent with the thermal conductivity measurement 
for X=ClO$_4$ suggesting a fully gapped state.\cite{BB97}
However, Shimahara has argued that a fully gapped state is possible even 
for $d$-wave pairing particularly in (TMTSF)$_2$ClO$_4$ , because in this case,
anion ordering takes place above the superconducting $T_c$, so that 
a ``gap'' opens up on the Fermi surface at positions $(k_a=\pm\pi/4)$ 
where the nodes of the superconducting gap would otherwise intersect 
(see Fig.\ref{fig2}(a)).\cite{Shimahara2}
Exactly the same argument holds for $f$-wave pairing since the positions  
of the gap nodes on the Fermi surface are the same between $f$ and $d$.
A fully gapped state usually results in a coherence peak followed 
by an exponential decay in $1/T_1$ as mentioned in 
section \ref{exptmtsf} \cite{HaseFuku}, 
which seems to be 
in contradiction with refs.\citen{Takigawa} and \citen{Lee3}, 
but since $1/T_1$ can be affected by the presence of impurities\cite{Hotta}, 
vortices\cite{TakiIchi}, or correlation effects\cite{Fujimoto},
the clarification of the relation between  $1/T_1$ and 
the thermal conductivity 
experiments is open for future study.

From the microscopic view discussed in the preceding sections, 
$f$-wave and $p$-wave are the main candidates 
for spin triplet pairing.\cite{Lee1,Lee2,Lee3}
As for the direction of $\Vec{d}$, 
if we assume that the spin fluctuations in the $\Vec{c}$ direction are  
weak while those in the $a$-$b$ planes have similar magnitude, 
$\Vec{d}$ of $f$-wave pairing should lie in the 
$\Vec{c}$ direction as discussed in section \ref{fwavetmtsf}, which is  
consistent with the Knight shift results.\cite{Lee1,Lee2} 
If we assume on the other hand that the spin fluctuations 
are solely strong in the $\Vec{b}'$ direction 
compared to those in the $a$-$c$ plane, then 
 $f$-wave's $\Vec{d}$ is perpendicular to $\Vec{b}'$ and 
lies in the $a$-$c$ plane as also discussed in section \ref{fwavetmtsf}, 
which is more closer to the $\Vec{d}$
direction proposed by Lebed {\it et al.}\cite{LMO} 
from the temperature dependence 
of $H^b_{c2}$ and $H^a_{c2}$.\cite{Lee2}
In the case of $p$-wave pairing, 
if the anisotropic spin fluctuations 
contribute to the pairing interaction 
in the form given in eqs.(\ref{phononspin}) 
($V_{\rm el-ph}$ need not be due to 
phonons), $\Vec{d}$ is likely to lie  
in the $a$-$b$ plane. Thus, it may be possible to distinguish $f$ and $p$
from the direction of $\Vec{d}$, provided that the 
anisotropic spin fluctuations play a role in the Cooper pairing.
Such a test, however, has to be done in the absence of, or under low,
magnetic field since $\Vec{d}$ may rotate   
regardless of the pairing symmetry 
if the  magnetic field is sufficiently large 
to overcome the effect of the magnetic anisotropy. 

Although there exist few experiments up to date,
possibility of determining the pairing symmetry from 
tunneling spectroscopy measurements has been proposed theoretically 
by several groups.
Sengupta {\it et al.} pointed out that the presence/absence of 
zero energy peak in the tunneling conductance 
can be used to distinguish 
various types of pairings in (TMTSF)$_2$X.\cite{Sengupta} 
In fact, the zero energy peak in the tunneling spectroscopies 
of anisotropic superconductors (those with 
sign change in the gap) originates from the zero-energy 
Andreev bound state caused 
by the sign change of the pair potential felt by the quasiparticle 
in the reflection process at the surface,\cite{Hu,TanaKash}
and has turned out to be a powerful method for probing the 
pairing symmetry in anisotropic superconductors such as  
the high $T_c$ cuprates.\cite{TanaKashRev}
In the case of tunneling parallel to $\Vec{a}$  
in particular, the zero energy 
peak does not exist for $d$-wave. This is because 
the injected and the reflected quasiparticles feel the same gap
due to  $\Delta_d(k_a,k_b)=\Delta_d(-k_a,k_b)$ (see Fig.\ref{fig2}(b)).
By contrast, the zero energy peak does exist for $p$-wave and $f$-wave,
where $\Delta_{f,p}(k_a,k_b)=-\Delta_{f,p}(-k_a,k_b)$ 
is satisfied (Fig.\ref{fig2}(c)).
Tanuma {\it et al.} further pointed out that $p$-wave and $f$-wave 
can be distinguished from the overall shape of the surface density of 
states (overall structure of the tunneling spectrum) because 
$p$-wave is a fully gapped state, which results in a U-shaped surface
density of states around the Fermi level, 
while $f$-wave results in a V-shaped one.\cite{Tanuma}
Therefore, the combination of the absence/presence of the 
zero energy peak and the overall shape of the spectrum 
enables us to distinguish  $p$, $d$, and $f$-wave pairings.

Further theoretical studies based on various shapes  
of the Fermi surface have been performed.\cite{Tanuma2,Asano}
Tanuma {\it et al.} showed that when the Fermi 
surface is warped in a certain manner, the zero energy peak can 
appear even in the case of $d$-wave. In this case, $d$ and $f$-wave 
can be distinguished by the way the zero energy peak splits 
in the presence of a magnetic field.\cite{Tanuma2} 
Such studies show that the existence of the 
zero energy peak is sensitive to  
the shape of the Fermi surface (compare Fig.\ref{fig2}(a)and (d), or 
(b) and (e)). Since the hopping integrals, and thus the Fermi surface, 
of (TMTSF)$_2$X depend on the 
pressure, the temperature, and the anions\cite{Grant,Ducasse,commentband},  
it is necessary to 
strictly pin down the actual shape of the Fermi surface 
at the temperature and the pressure at which superconductivity takes place 
in order to distinguish the pairing symmetry from the 
presence/absence of the zero energy peak.

The tunneling tests above mainly concern the orbital part of the pairing.
On the other hand, Bolech and Giamarchi proposed a tunneling experiment to 
distinguish directly the spin part of the pairing.\cite{Bolech} 
They showed that the $I$-$V$ 
characteristics of a normal metal-triplet superconductor junction 
are unaffected by an application of magnetic field 
perpendicular to $\Vec{d}$, while the Zeeman effect affects 
the $I$-$V$ characteristics when $\Vec{d}\parallel\Vec{H}$ 
similarly to the case of normal metal-singlet superconductor junction.
Therefore,  the spin part of the pairing,
whether it is singlet or triplet and also the direction of $\Vec{d}$ if 
triplet, can be determined 
by measuring the $I$-$V$ characteristics of the junction 
under a rotating magnetic field, provided that 
$\Vec{d}$ does not rotate according to the direction of the 
magnetic field. 
Vaccarella {\it et al.} also proposed a way of directly probing the 
spin part of the triplet pairing. Namely, they showed that 
the Josephson effect between 
two triplet superconductors is very sensitive to the direction 
of $\Vec{d}$ across the junction, and proposed that 
this sensitivity can be used as a test for triplet superconductivity.
\cite{Vaccarella}

As a final remark in this subsection, it is important to recognize 
that the pairing symmetry might be different for different anions.
This possibility is suggested especially from the viewpoint discussed 
in section \ref{fwavetmtsf}. For instance, the amplitude of 
$2k_F$ charge fluctuations, which has to be large 
for $f$-wave to dominate over $d$-wave, is found to be small for 
X=AsF$_6$, so that $f$-wave has less chance of dominating over 
$d$-wave than in the case of X=PF$_6$. Thus, 
the pairing symmetry of a TMTSF superconductor with a certain anion 
should be determined by a combination of multiple experiments on 
the salt with that very anion. Furthermore, we must keep in mind that 
the pairing symmetry might even change for the same salt 
under different environment, such as the pressure and the 
strength of the magnetic field, because several pairing symmetries 
may be closely competing. In particular, as mentioned in 
section \ref{fwavetmtsf}, 
singlet to triplet transition may take place under high magnetic field
since the singlet pairing is Pauli-paramagnetically 
limited.\cite{Fuseya04,Shimahara6,Vaccarella2}

\section{$\kappa$-(BEDT-TTF)$_2$X}
\subsection{Lattice Structure and the Phase Diagram}
\label{latticebedt}
\begin{figure}
\begin{center}
\includegraphics[width=8cm,clip]{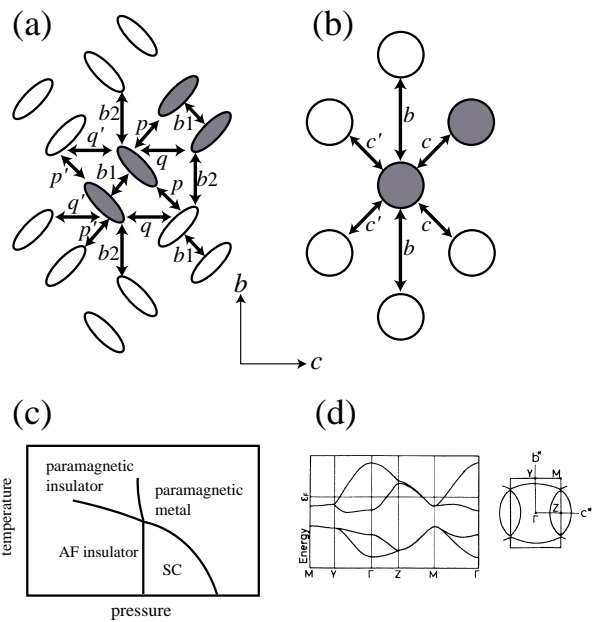}
\end{center}
\caption{(a) The lattice structure of $\kappa$-(BEDT-TTF)$_2$X in the 
$b$-$c$ plane. $b_1$, $b_2$, $\cdots$ represent the hopping integrals in
 the four band model. (b) The lattice structure of the dimer model.
(c) Phase diagram of $\kappa$-(BEDT-TTF)$_2$X.\cite{kappaRev} 
(d) Band structure 
and the Fermi surface of $\kappa$-(BEDT-TTF)$_2$Cu(NCS)$_2$.
(Reprinted with permission from 
ref.\citen{Oshima}. Copyright 1988 by the American Physical Society.)}
\label{fig6}
\end{figure}

The lattice structure of $\kappa$-(BEDT-TTF)$_2$X is shown in 
Fig.\ref{fig6}(a), which consists of dimers formed by a pair of 
face-to-face molecules. $b$- and $c$-axis are taken as in 
Fig.\ref{fig6}, while the BEDT-TTF layers and the anion layers 
alternate along the $a$-axis. 
Relatively large overlap of the orbitals between 
the dimers exists (see section \ref{elecstrucbedt}) 
while the overlap between the 
BEDT-TTF layers is very small, resulting in a strong 
two dimensionality.

In Fig.\ref{fig6}(c), the generic phase diagram of $\kappa$-(BEDT-TTF)$_2$X is
shown, which has been extensively studied by 
Kanoda {\it et al.}\cite{kappaRev} The superconducting and 
the antiferromagnetic insulating phases share a first order phase boundary.
Recently, this boundary has been revealed to persist 
above the superconducting $T_c$ and the 
N\'{e}el temperature into the boundary of the paramagnetic insulating  
and the metallic phases, ending up at a certain critical point
\cite{Fournier,Kagawa}, where an anomalous criticality has been 
found recently.\cite{Kagawa2} 
The horizontal axis in the phase diagram 
can be considered as hydrostatic or chemical pressure, 
where superconductivity with $T_c$ exceeding 10K occurs 
at ambient pressure for $\kappa$-(BEDT-TTF)$_2$Cu(NCS)$_2$
\cite{Urayama} and  
$\kappa$-(BEDT-TTF)$_2$Cu[N(CN)$_2$]Br, while 
$\kappa$-(BEDT-TTF)$_2$Cu[N(CN)$_2$]Cl is an antiferromagnetic insulator
below 26K at ambient pressure\cite{Miyagawa} and becomes superconducting with 
$T_c=12.8$K under an applied pressure of 0.3kbar.\cite{Wang}

\subsection{Electronic structure}
\label{elecstrucbedt}
\begin{figure}
\begin{center}
\includegraphics[width=8cm,clip]{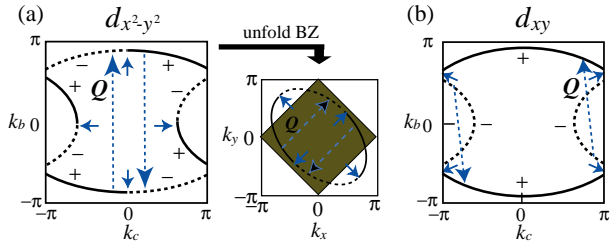}
\end{center}
\caption{(a) The $d_{x^2-y^2}$-wave gap in the original Brillouin
zone and in the unfolded one (right). Note that although the gap changes 
sign at the Brillouin zone edge $(k_c=\pm \pi)$, the nodes of the gap 
are not located there; the gap jumps from a positive to a negative value, 
as can be seen more clearly in the unfolded Brillouin zone. 
(b) The $d_{xy}$-wave gap. Here, we show the case when the two portions of the 
Fermi surface splits due to the lack of center-of-inversion symmetry. 
In this case, the $d_{xy}$ gap nodes do not intersect the Fermi surface,
although the gap does become small near the Brillouin zone edge.
On the other hand, if the two portions stick,
the $d_{xy}$ nodes intersect the Fermi surface at the Brillouin zone edge.
The solid (dashed) curves represent the portions of the Fermi surface 
where the gap has a positive (negative) sign. $\Vec{Q}$ represents 
the wave vector of the spin fluctuation mode that favors each pairing 
symmetry. The short arrows denote the positions of the gap nodes.
}
\label{fig7}
\end{figure}

The values of the intermolecular hopping integrals shown in 
Fig.\ref{fig6}(a) have been estimated using the extended H\"{u}ckel method,
\cite{Oshima,Komatsu} which is summarized in Table \ref{hoppingbedt}.
The hopping integral in the 
$b$-direction alternates as $t_{b1}, t_{b2}, t_{b1},\cdots$, 
where $|t_{b1}|>|t_{b2}|$ because of the dimerization of the molecules.
In Fig.\ref{fig6}(d), the band structure of 
$\kappa$-(BEDT-TTF)$_2$Cu(NCS)$_2$\cite{Oshima} is shown. 
Four bands exist near the Fermi level because there are four BEDT-TTF 
molecules per unit cell. Due to the dimerization, a gap 
opens up between the bonding and the antibonding bands.
There is one hole per dimer, so that only 
the upper two bands cross the Fermi level, resulting in 
two portions of the Fermi surface (Fig.\ref{fig7}).
The two portions of the Fermi surface are connected at the Brillouin
zone edge for X=Cu[N(CN)$_2$]Br, Cu$_2$(CN)$_3$, I$_3$, etc., in which the 
anions are arranged in a manner that the system possesses
center-of-inversion symmetry, which results in $t_p=t_p'$, $t_q=t_q'$.
On the other hand, for X=Cu(NCS)$_2$, the system 
lacks the symmetry so that $t_p\neq t_p'$, $t_q\neq t_q'$, 
and in that case, the two portions of the 
Fermi surface are separated into an open Fermi surface and a closed one.

In the limit of large $t_{b1}$, namely, when the dimerization is strong, 
each dimer can be considered as a single site, so 
the system reduces to a two band model shown in Fig.\ref{fig6}(b)
with $n=1$, where the band filling is now defined as 
$n$=(the number of electrons/the number of {\it sites}).\cite{Tamura,KF} 
In other words, the energy gap between the upper two and the lower
two bands becomes large when the dimerization is strong, so that the 
lower two bands, which do not cross the Fermi level, can be neglected. 
In this strong dimerization limit, 
the effective hopping integrals $t_b$ and $t_c$ are given 
as $t_b=-t_{b2}/2$, $t_c=(-t_p+t_q)/2$, and $t_c'=(-t_p'+t_q')/2$
in terms of the original hopping integrals,\cite{Tamura} which gives 
$|t_b/t_c|\sim 0.8$ for X=Cu(NCS)$_2$ and  
$|t_b/t_c|\sim 0.7$ for X=Cu[N(CN)$_2$]Br.
The system further reduces to a half-filled single band model when $t_c=t_c'$. 
\begin{table}
\begin{tabular}{lcccccc}
anion          &  $b_1$  &  $b_2$   & $p$    & $p'$   & $q$   & $q'$ \\
\hline 
Cu$_2$(CN)$_3$ & 22.36 & 11.54  & 8.01 & $-$ & $-2.90$ & $-$ \\
Cu(NCS)$_2$    & 22.95 & 11.31  & 9.85 & 10.09 & $-3.30$ & $-3.76$ \\
Cu[N(CN)$_2$]Br& 24.37 & 9.16   & 10.14 &  $-$  & $-3.40$ &  $-$ 
\end{tabular}
\caption{Hopping integrals estimated in ref.\citen{Komatsu}. In units of 
10$^{-2}$eV.}
\label{hoppingbedt}
\end{table}

\subsection{Experimental Results Concerning the Pairing Symmetry}
\label{expbedt}
Here, we summarize the experimental results concerning the 
pairing symmetry.\cite{kappaPairRev}
In the NMR experiments for $\kappa$-(BEDT-TTF)$_2$Cu[N(CN)$_2$]Br,
the $^{13}$C Knight shift has been found to decrease below
$T_c$,\cite{Mayaffre,Soto} which is consistent with singlet pairing.
Also, the $^{13}$C spin-lattice relaxation rate $1/T_1$ 
for $\kappa$-(BEDT-TTF)$_2$Cu[N(CN)$_2$]Br exhibits no coherence 
peak, and a power law decay proportional to $T^3$ is seen below $T_c$.
\cite{Mayaffre,Soto,Kanoda1}
As in the case of (TMTSF)$_2$X, this is consistent with the 
presence of line nodes in the superconducting gap.
Also, in a thermal conductivity measurement for X=Cu(NCS)$_2$, 
a $T$-linear term has been found at low temperatures, suggesting 
the existence of nodes in the gap.\cite{Belin}

On the other hand, there has been much controversy concerning the 
measurements of other quantities.
The magnetic penetration depth has been measured using techniques such as 
muon spin relaxation,\cite{Harshman,Le} 
ac susceptibility,\cite{Kanoda2,Pinteric}
surface impedance,\cite{Dressel,Achkir} and  
dc magnetization.\cite{Lang,Lang2}.
The penetration depth should exhibit an exponentially decaying 
behavior for a fully gapped state, while a power-law dependence is 
expected at low temperatures for gaps with nodes.
Some studies have found for X=Cu(NCS)$_2$ 
that the temperature dependence of the 
penetration depth is consistent with a conventional full gap state, 
\cite{Lang,Dressel,Harshman} while others have found results
consistent with a gap with nodes.
\cite{Kanoda2,Le,Achkir,Carrington}
Similar controversy on the penetration 
depth also exists for X=Cu[N(CN)$_2$]Br, 
where the presence of gap nodes
\cite{Le,Carrington,Pinteric} as well as the absence of 
them\cite{Lang,Dressel,Lang2} has been suggested.

The temperature dependence of the specific heat has also been another 
issue of controversy. Nakazawa and Kanoda found for X=Cu[N(CN)$_2$]Br
a $T^2$ dependence of the 
electronic specific heat,\cite{Kanoda3} 
which was taken as an indication for the presence of nodes in the 
gap. However, more recent results for X=Cu[N(CN)$_2$]Br\cite{Elsinger}
and for X=Cu(NCS)$_2$\cite{Muller} have shown 
exponentially activated temperature 
dependence, indicating a fully gapped superconducting state.

The above experiments 
do not give direct information on the position of, if any,  
the nodes in the gap function. Several groups have in fact made attempts to 
directly determine the node positions.
A millimeter-wave transmission experiment suggested a 
gap function which has nodes in the direction shown in Fig.\ref{fig7}(a).
\cite{Schrama} 
If we unfold the Brillouin zone (right panel of Fig.\ref{fig7}(a)), 
which corresponds to adopting a single dimer as a unit cell, 
(this is possible when $t_c=t_c'$), 
this gap function has nodes in the diagonal direction 
like the gap in the high $T_c$ cuprates.\cite{SigristUeda,TanaKashRev}
In this sense, we will refer to this pairing symmetry as 
$d_{x^2-y^2}$-like pairing hereafter.
(Note that this terminology is the opposite to the one adopted in 
ref.\citen{KKATM}. Namely, we define the $x$ and $y$ axes 
by rotating $b$ and $c$ axes by 45 degrees.)
However, different interpretations on this experiment have been 
proposed afterwards.\cite{Hill,Shibauchi}
On the other hand, Arai {\it et al.} showed 
for $\kappa$-(BEDT-TTF)$_2$Cu(NCS)$_2$ 
using in-plane STM measurement 
that the gap is the largest in the $b$ and $c$ directions. 
\cite{Arai} This is more consistent with the gap function shown in 
Fig.\ref{fig7}(b), which has nodes in between the $k_b$ and $k_c$ directions.
Izawa {\it et al.} measured the thermal conductivity of 
$\kappa$-(BEDT-TTF)$_2$Cu(NCS)$_2$ under a magnetic field rotating 
in the $b$-$c$ plane, where an oscillation again consistent with 
the gap in Fig.\ref{fig7} (b) was observed.\cite{IM} 
This conclusion is
based on theoretical studies\cite{Vekhter} showing that 
the density of states of a superconducting state having gap nodes in 
some direction becomes large (small) when the magnetic field is applied in the 
antinodal (nodal) direction due to the Doppler shift of the quasiparticle 
energy spectrum.\cite{Volovik}
Since the gap in Fig.\ref{fig7}(b) 
has nodes in the vertical and horizontal directions 
in the unfolded Brillouin zone, we will call this gap  
$d_{xy}$-like hereafter.

\subsection{Phonon Mechanisms}
\label{phononbedt}
If the symmetry is $s$-wave, the pairing 
is most likely due to electron-phonon interactions.
Here we give a brief survey on the electron-phonon-interaction mechanism 
proposed for $\kappa$-(BEDT-TTF)$_2$X. 
Yamaji argued that intramolecular phonons (molecular vibrations)
should play an important role in the occurrence of superconductivity 
in organic materials. The theory in which the coupling between the 
electrons and the molecular vibrations is taken into account 
was applied to $\beta$-(BEDT-TTF)$_2$I$_3$\cite{Yamaji4}
and to $\kappa$-(BEDT-TTF)$_2$Cu(NCS)$_2$\cite{ISY}, where $T_c$ for $s$-wave 
superconductivity was estimated.

Girlando {\it et al.}\cite{Girlando} 
considered the coupling between the electrons and both 
the intramolecular and the 
intermolecular phonons (lattice vibrations), and estimated the $s$-wave $T_c$ 
for $\kappa$-(BEDT-TTF)$_2$I$_3$ and $\beta$-(BEDT-TTF)$_2$I$_3$
using the Allen-Dynes formula.\cite{AllenDynes}
They concluded that the contributions of both phonons are important 
in understanding the experimental values of the $T_c$.

On the other hand, 
Varelogiannis considered the phonons with small $\Vec{q}$ due to 
weak screening,\cite{Varelogiannis} 
similar to those discussed in section \ref{phonontmtsf} for TMTSF salts.
Adopting the single band dimer model, 
the BCS gap equation was solved by 
assuming the phonon-mediated 
attractive interaction in the form $\propto -1/(q_c^2+\Vec{q}^2)$, where 
$q_c$ is a momentum cutoff parameter, and also taking into account the 
Coulomb pseudopotential. There, it has been found that 
a close competition between an anisotropic 
$s$-wave  and $d_{x^2-y^2}$-wave  pairings takes place. 

\subsection{Studies on the dimer model}
\label{dimerbedt}
If the superconducting gap in $\kappa$-(BEDT-TTF)$_2$X indeed 
has nodes, the most probable scenario is that
electron correlation plays an important role in the pairing.
In this and the next subsection, we discuss electronic 
mechanisms of superconductivity having gap with nodes.

Following the study by Kino and Fukuyama, who showed that the 
insulating state in $\kappa$-(BEDT-TTF)$_2$X can be 
understood as a Mott insulator of the half-filled dimer model in which 
the on-site repulsion $U$ (which corresponds to the effective 
repulsion within the dimer) is considered,
\cite{KF,kappaRev}
various theoretical studies concerning the superconductivity 
have been performed for the dimer model with on-site $U$ 
on the two band\cite{Schmalian}(for $t_c\neq t_c'$) 
or the single band (for $t_c=t_c'$) lattice 
\cite{KinoKontani2,KondoMoriya,KA,Dagotto,Louati,Jujo,Liu} 
with $t_b/t_c= 0.6\sim 0.8$.
In these studies, 
the $d_{x^2-y^2}$-wave has been found to be the most dominant pairing.
Namely, in RPA or FLEX studies,
\cite{Schmalian,Dagotto,KondoMoriya,KinoKontani2,Louati}
the spin susceptibility is found to have a peak at the nesting 
vector $\Vec{Q}$ (near $(\pi,\pi)$ 
in the unfolded Brillouin zone, see Fig.\ref{fig7}(a)) 
that bridges the open portions of the Fermi surface, 
although the nesting is not so good. Then, in order to 
have opposite signs of the gap across $\Vec{Q}$, and also to satisfy the 
even parity condition for spin-singlet pairing, 
the $d_{x^2-y^2}$ gap is favored. Within the FLEX studies, $T_c$ has 
been estimated to be O(10K),\cite{KondoMoriya,KinoKontani2} 
consistent with the experiments.
By using the third order perturbation theory, Jujo {\it et al.} 
showed that $d_{x^2-y^2}$-wave pairing dominates but with  $T_c$ 
lower than those obtained in FLEX\cite{Jujo}. They concluded 
that the vertex corrections that are not taken into account in 
FLEX have an effect of suppressing the $T_c$
especially for systems on frustrated lattices, at least up to third 
order. As for numerical approaches for finite size systems,
Kuroki and Aoki applied the ground state QMC technique 
and showed that the $d_{x^2-y^2}$-wave  pairing correlation function is 
enhanced at large distances accompanied by a development of the 
spin correlation near $\Vec{q}=(\pi,\pi)$.\cite{KA} 
Quite recently, Liu {\it et al.} used the variational Monte Carlo 
technique, where $d_{x^2-y^2}$-wave superconducting order 
parameter is found to be enhanced in a certain range of $U/t$.\cite{Liu}

There have also been some strong coupling approaches along the 
line of Anderson's resonating valence bond (RVB) theory for the 
high $T_c$ cuprates,\cite{Anderson} in which large $U/t$ is assumed. 
In the large $U/t$ limit, the Hubbard model on a square lattice 
is transformed into the 
$t$-$J$ model which consists of the nearest neighbor antiferromagnetic 
superexchange term (the magnitude of the superexchange being 
$J=4t^2/U$, where $t$ is the nearest neighbor hopping) 
and the hopping term in the space that prohibits double occupancy of 
electrons at a single site.
At exactly half filling, the hopping term vanishes, so that the 
model reduces to the antiferromagnetic Heisenberg model, which 
describes the experimental situation for the cuprates to be an 
antiferromagnetic Mott insulator when carriers are not doped, i.e., 
for the half-filled band.\cite{revtJ}  However, 
in $\kappa$-(BEDT-TTF)$_2$X, a difference from the cuprates lies in that 
the band filling (of the dimer model) remains at half filling even 
when metallized or superconducting upon increasing the (chemical) pressure. 
In this context, Baskaran pointed out the 
possibility of ``self-doping'' of carriers at half filling, where 
an equal number ($N_0$) of doubly occupied sites and empty sites hop in the 
background of singly occupied sites, which is shown to be equivalent to the 
usual $t$-$J$ model with $2N_0$ holes.\cite{Baskaran} 
On the other hand, Powell and McKenzie studied a model that contains the 
nearest neighbor (which will be denoted as $J_c$ here) 
and the next nearest neighbor ($J_b$) 
superexchange terms in addition to the Hubbard model.\cite{PowellMck2} 
In this model, the double occupancy of a site is not prohibited, so that 
the hopping term does not vanish at half filling, thereby circumventing 
the difficulty in the usual $t$-$J$ model, 
 which is always insulating at half filling.
Assuming the parameter values $J_c/t_c=1/3$ and $J_b/J_c=(t_b/t_c)^2<1$,
a first order transition from $d_{x^2-y^2}$-wave superconductivity to 
a Mott insulating state was shown to occur upon increasing $U$ 
within the Hartree-Fock-Gor'kov approximation.

\subsection{Studies for the Hubbard model on the original four band lattice}
\label{fourbandbedt}
The $d_{x^2-y^2}$ pairing 
in the dimer model for $t_b/t_c<1$\cite{KondoMoriya3} nevertheless contradicts 
at least with the experimental observation 
of $d_{xy}$-like gap for X=Cu(NCS)$_2$ in 
the thermal conductivity\cite{IM} and STM experiments.
\cite{Arai}
Motivated by this discrepancy between the theories and the experiments, 
Kuroki {\it et al.}\cite{KKATM} performed a FLEX study on 
the original four band lattice (Fig.\ref{fig6}(a)), where 
they adopted the hopping integral values obtained from extended 
H\"{u}ckel calculation for X=Cu(NCS)$_2$ given in
Table \ref{hoppingbedt}.\cite{Komatsu} 
It can be seen from these hopping integral 
values that the ratio between the intradimer hopping ($t_{b1}$) and the 
largest interdimer ones ($t_{b2}$) is about 2,\cite{commentXu}  
which may not be considered as so large.

\begin{figure}
\begin{center}
\includegraphics[width=8cm,clip]{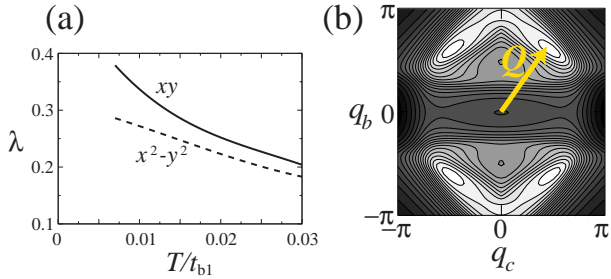}
\end{center}
\caption{The FLEX calculation results 
on the original four band lattice.\cite{KKATM} 
The adopted hopping integrals are those for 
X=Cu(NCS)$_2$. (a)Eigenvalues of the \'{E}liashberg equation for 
$d_{xy}$ and $d_{x^2-y^2}$ pairings plotted as functions of
 temperature. (b)Contour plot of the spin susceptibility at $T/t_{b1}=0.01$.
}
\label{fig8}
\end{figure}

From the temperature dependence of the 
 eigenvalues of the linearized \'{E}liashberg equation for the 
two types of  pairing shown in Fig.\ref{fig8}(a),  they concluded that  
 $d_{xy}$-like pairing dominates over $d_{x^2-y^2}$.
The origin of this result can be found in the spin structure.
Namely, the spin susceptibility $\chi$, shown in Fig.\ref{fig8}(b) 
peaks around ${\bf Q}\sim (\pm 0.4\pi,\pm 0.6\pi)$,
which is a consequence of a partial nesting between the open and the closed 
portions of the Fermi surface (Fig.\ref{fig7}(b)).
As a result, the gap function changes sign between 
the two portions of the Fermi surface, but does not change 
sign within each portion.

When the dimerization is strong, the four band model 
approaches the single band model, so that the nesting shown 
in Fig.\ref{fig7}(a) dominates and 
$d_{xy}$-like pairing gives way to $d_{x^2-y^2}$
when $t_{b1}$ is large.\cite{KKATM}
In fact, further four band analysis has shown that whether 
$d_{xy}$ or $d_{x^2-y^2}$ dominates depends on (a) the strength of the 
dimerization, (b) whether $t_b/t_c[=-t_{b2}/(-t_p+t_q)]$ 
is close to unity (which is a measure 
for how close the system is to an isotropic triangular lattice in the 
strong dimerization limit), 
and (c) the magnitude of the Fermi surface splitting due to 
$t_c\neq t_c'$ (i.e., $t_p\neq t_p'$, $t_q\neq t_q'$).\cite{KuroTana2} 
Namely, $d_{xy}$ is more favored when 
the dimerization is weaker, $t_b/t_c$ is closer to unity (closer to 
isotropic triangular lattice), and the Fermi surface splitting is
larger. These factors strongly 
depend on the anions. For X=Cu[N(CN)$_2$]Br, 
the dimerization is stronger 
($t_{b1}/t_{b2}\sim 2.7$ as compared to $\sim$2.0 for Cu(NCS)$_2$), 
$t_b/t_c\sim 0.7$ is away from unity   
(compared to $\sim$0.8 
for Cu(NCS)$_2$, see Table \ref{hoppingbedt}),
and there is no Fermi surface splitting because $t_c=t_c'$. 
In this case, $d_{x^2-y^2}$-wave is expected to dominate over $d_{xy}$ 
at least within the FLEX approach.


Although the results of the four band approach seem to be consistent 
with the $d_{xy}$-wave pairing observed in the thermal conductivity
\cite{IM} and the STM experiments\cite{Arai} for 
$\kappa$-(BEDT-TTF)$_2$Cu(NCS)$_2$, 
there remains an issue concerning the $T_c$.
Namely, $T_c$ exceeding 10K for X=Cu(NCS)$_2$ 
is almost the highest among charge-transfer-type organic superconductors.
On the other hand, the eigenvalue $\lambda$ in Fig.\ref{fig8} 
remains small in the temperature range studied in the FLEX study.
Although the low temperature regime was not studied in ref.\citen{KKATM} 
due to the restriction of the calculation, it is questionable whether the 
eigenvalue actually reaches unity at around $T\sim 0.004t_{b1}$, which 
corresponds to the actual $T_c\sim 10$K. The eigenvalue becomes 
larger in the presence of Fermi surface splitting (which was not 
considered in ref.\citen{KKATM}) because the 
nodes of the $d_{xy}$ gap do not intersect the Fermi surface,
but the $T_c$ problem still exists even in that case.\cite{KuroTana2} 
In fact, Kondo and Moriya have studied the four band model 
using FLEX, and showed that a realistic $T_c$ can be 
obtained only when the dimerization is extremely strong 
($|t_{b1}/t_{b2}|>5$).\cite{KondoMoriya2}. 
Such a strong dimerization indeed contradicts with 
the extended H\"{u}ckel\cite{Oshima,Komatsu} or 
the first principles calculations,\cite{Xu,Ching}
and also leads to the $d_{x^2-y^2}$-wave pairing, which  
is not in agreement at least with the thermal conductivity\cite{IM} 
and the STM experiments\cite{Arai} for X=Cu(NCS)$_2$.

This problem may be due to one (or more) of the following 
possibilities.
(i) The FLEX approximation is not sufficient for quantitative estimation of 
$T_c$ particularly for the $\kappa$-type BEDT-TTF salts 
because the nesting of the Fermi surface is not good, so that 
a strong development of a single spin fluctuation mode, 
which is necessary to verify RPA-like 
approaches, is absent especially when the dimerization is not so strong. 
Note that the spin fluctuations tend to be weak when 
the dimerization is weak because the band 
becomes more closer to 3/4 filling, where the effect of the 
on-site repulsion $U$ is weaker than for half filling.
(ii) In the limit of weak dimerization, the band is 3/4-filled and 
thus far away from half filling.
In systems away from half filling, terms neglected in the FLEX approximation, 
such as the vertex corrections, may play an important role.
(iii) The Hubbard model 
is oversimplified, and additional terms such as the off-site (inter-molecular)
repulsions and/or electron-phonon interactions are necessary.
Note that a phonon-mediated attractive interaction that gives 
large contribution around $\Vec{q}\sim 0$ as considered in 
section \ref{phononbedt} 
is likely to enhance all types of pairings, so that a consideration of 
electron-phonon interaction together with the 
spin fluctuations might result in 
a superconductivity with appropriate $T_c$, 
maintaining the dominant gap symmetry obtained by  
applying FLEX to the purely repulsive model.

Whether the above (i)$\sim$(iii) 
actually provides solution to this $T_c$ puzzle 
remains  open for future study,\cite{commentFLEX} 
but some hints have been found in 
the recent theoretical studies on $\beta'$-(BEDT-TTF)$_2$ICl$_2$,
\cite{KKM,NakanoKuroki,Seo} a superconductor 
which has been found under high pressure by Taniguchi {\it et al}.
\cite{Taniguchi}
$\beta'$-(BEDT-TTF)$_2$ICl$_2$ also has dimerization of molecules,
and the $T_c(\sim 14$ K) is the highest among the 
charge-transfer type molecular solids and somewhat close to the $T_c$ 
of the $\kappa$-type salts, so that the comparison between the 
two types is intriguing.
Using the hopping integral values obtained from 
the first principles calculation by Miyazaki and Kino,\cite{MiyaKino}
Kino {\it et al.} performed a FLEX study on the single band dimer model, and 
obtained a phase diagram in the pressure-temperature space, which 
is similar to the experimental phase diagram including the values of $T_c$,
although the superconducting phase is shifted to a somewhat higher 
pressure regime.\cite{KKM} The superconducting gap in this case has a 
$d_{xy}$-like structure, reflecting the good nesting of the Fermi surface.
Later, Nakano and Kuroki\cite{NakanoKuroki} 
performed 
a FLEX study on the two-band lattice with finite dimerization.   
This was motivated by the fact that the ratio between the intradimer 
hopping integral and the largest interdimer one 
was estimated to be about $\sim 2$,\cite{MiyaKino} which is 
a situation similar to the 
case of $\kappa$-(BEDT-TTF)$_2$Cu(NCS)$_2$\cite{Komatsu}, so that 
adopting the dimer model might be questionable. 
Nevertheless, the gap structure, $T_c$, and the phase diagram 
obtained in the two-band approach is found to be quite 
similar to the one obtained in the dimer model
approach.\cite{NakanoKuroki} 
This suggests that the possibility (ii) mentioned above is not, or 
at least not always, the case. The fact that the Fermi surface of 
$\beta'$-(BEDT-TTF)$_2$ICl$_2$ is well nested may have some relevance to the 
point (i) mentioned above, as pointed out in ref.\citen{NakanoKuroki}.

\subsection{Tests for the Pairing Symmetry Candidates}
\label{testbedt}
Here we survey some theoretical 
proposals for further experimental 
tests on the pairing symmetry.\cite{commenttest}
Powell and McKenzie\cite{PowellMck} showed that the $T_c$ dependence on the 
level of structural disorder\cite{Su,Stalcup} can be 
explained by assuming $d$-wave pairing and using the 
formula (\ref{AGform}). Since the same applies to 
$s$-wave pairing in the presence of magnetic impurities, 
and there is indeed a possibility that disorder leads to a formation 
of local magnetic moments due to the proximity to antiferromagnetism, 
the two possibilities ($d$+non-magnetic defects or $s$+magnetic ones) 
cannot be distinguished within these experiments
alone. They further proposed probing the presence/absence of 
magnetic impurities by, e.g., a $\mu$SR experiment.

Li showed that $d_{xy}$, $d_{x^2-y^2}$, and $s$-wave 
superconducting states all exhibit different spin structures, 
so that they can be distinguished by looking at the spin 
susceptibility below $T_c$.\cite{Li} Namely, 
for $s$-wave, a full gap opens 
on the Fermi surface so that the spin susceptibility is strongly suppressed.   
As for the pairings with gap nodes, 
since $d_{xy}$ and $d_{x^2-y^2}$-wave pairings open up a gap 
at different portions of the Fermi surface, different types of 
nesting take place (as mentioned in section \ref{fourbandbedt}, two types 
of nesting are possible for the Fermi surface of the $\kappa$ salts),
resulting in a different spin structure. 

Tanuma {\it et al.} proposed a magnetotunneling spectroscopy 
for determining the pairing symmetry.\cite{Tanuma5} Namely, as mentioned 
in section \ref{expbedt}, the density of states of a superconductor having 
nodes in the gap oscillates by rotating the direction of the 
magnetic field.\cite{Vekhter} 
This oscillation can be detected by looking 
at the surface density of states in the tunneling spectroscopy.
The phase of the oscillation is different between 
$d_{x^2-y^2}$ and $d_{xy}$ pairings, so that the pairing symmetry 
can be determined as in the case of the thermal conductivity 
measurement.\cite{IM}

As in the case of TMTSF salts, it is worth mentioning that the 
pairing symmetry need not be the same for different anions 
since here again, the competition among the pairing symmetries 
may be close. 
As mentioned in section \ref{fourbandbedt}, 
the competition between $d_{xy}$ and 
$d_{x^2-y^2}$ can be affected by various factors, such as the 
splitting of the Fermi surface and the strength of the dimerization, 
which depend on the anions. 
Speaking of anion dependence, it is also important to notice that 
$d_{xy}$-wave pairing in $\kappa$-(BEDT-TTF)$_2$Cu(NCS)$_2$ may look 
like $s$-wave pairing from those experiments that detect only the 
amplitude of the gap because the nodes 
of the gap do not intersect the Fermi surface due to the 
splitting of the Fermi surface (see Fig.\ref{fig7}(b)) 
just like $d$-wave and  $f$-wave in the anion ordered (TMTSF)$_2$ClO$_4$. 

\section{Concluding Remarks}
\label{conclude}
In this paper, a review has been given on the theoretical studies 
concerning the pairing symmetry competition in (TMTSF)$_2$X and
$\kappa$-(BEDT-TTF)$_2$X. Some of the pairing symmetry candidates and 
their possible microscopic origins have been discussed. 
Existing experimental results concerning the pairing symmetry 
as well as some proposals for further tests have also been surveyed.
Close competition among 
different pairing symmetries makes the problem of theoretically 
pinning down the symmetry difficult, and at the same time, very
intriguing. One of the origin of this close competition is the 
peculiarity of the Fermi surface, i.e., disconnected Fermi surfaces in both 
(TMTSF)$_2$X and $\kappa$-(BEDT-TTF)$_2$X  (although in different
senses), and two types of nesting 
in $\kappa$-(BEDT-TTF)$_2$X.
This also implies that several pairing symmetries  
can be closely competing in the actual materials, so that the 
symmetry might be different for different anions, or 
even for the same anion under different environment 
(magnetic field, pressure, etc.).
Although a lot has been understood concerning the superconductivity 
in these materials, further theoretical and experimental 
studies are required for a more clear understanding of the nature of the 
superconducting state.

\acknowledgement
The author would like to thank
 H. Aoki, Y. Tanaka, R. Arita, Y. Matsuda, Y. Tanuma, 
T. Kimura, S. Onari, Y. Asano, S. Kashiwaya, M. Takigawa, M. Ichioka, 
T. Nakano, and H. Aizawa for collaboration 
on the theory of organic superconductors. 
He  is also grateful to H. Fukuyama, M. Ogata, J. Suzumura, K. Kanoda, 
G. Baskaran, J. Singleton, H. Seo, A. Kobayashi, 
Y. Ohta, Y. Fuseya, T. Takimoto, H. Kontani, T. Miyazaki, M. Kohmoto, 
and M. Sato for fruitful discussions.
He acknowledges 
Grants-in-Aid for Scientific Research from the Ministry of Education, 
Culture, Sports, Science and Technology of Japan, and from the Japan 
Society for the Promotion of Science.



\begin{thebibliography}{99}
\bibitem{symmetrycomment} The symmetries of $p$-wave and $f$-wave gaps  
are the same in the group theory sense. They both have the gap 
essentially of the form $\sin(mk_a)$, where $m$ is an integer. 
Thus, in section \ref{tmtsf}, we use the word `symmetry' 
in a more loose sense of the term.
\bibitem{commentmaganiso} All microscopic theories for the $d$-$f$ 
competition so far, to our knowledge, have neglected the 
anisotropy in the spin fluctuations.
\bibitem{commentgology} In $g$-ology,  the ``phase diagram'' is determined 
by comparing the correlation functions of various order parameters because 
there is no true long range order in pure one dimension. If the 
pairing correlation decays more slowly than other correlations, 
the phase is ``superconducting''.
\bibitem{commentband} Also, the hopping integrals estimated theoretically 
rather strongly depend on the adopted methods themselves, so a 
``smoking gun'' in this sense may be to directly determine the Fermi surface 
experimentally.
\bibitem{commentXu} It should further be noted that the dimerization 
gap between the bonding and the antibonding bands 
estimated from a first principles calculation\cite{Xu} 
is even smaller (O(0.01)eV) than that obtained from the 
extended H\"{u}ckel calculations (O(0.1)eV).
\bibitem{commenttest} In principle, experiments that have been performed for
the high $T_c$ cuprates and other possible unconventional
superconductors can be applied also to $\kappa$-(BEDT-TTF)$_2$X.
For reviews on these pairing symmetry tests, see e.g., 
refs.\citen{SigristUeda} and \citen{TanaKashRev}. Here we pick up some 
proposals that have been raised for $\kappa$-(BEDT-TTF)$_2$X in particular.
\end{thebibliography}
\end{document}